\documentclass[a4paper,10pt,superscriptaddress,aps,nofootinbib]{revtex4-1}
\pdfoutput=1
\usepackage{amsfonts,amssymb,amsmath,subfig,multirow}
\usepackage[pdftex]{graphicx,color}
\newcommand{\bra}[1]{\langle #1|}
\newcommand{\ket}[1]{|#1\rangle}
\newcommand{\braket}[2]{\langle #1|#2\rangle}
\newcommand{\im}{\textrm{Im}\ }

\newcommand{\beq}{\begin{equation}}
\newcommand{\eeq}{\end{equation}}
\newcommand{\ba}{\begin{array}{ccc}}
\newcommand{\ea}{\end{array}}
\newcommand{\nn}{\nonumber \\}

\newcommand{\vbs}{\ket{\text{VBS}}}
\def\bea{\begin{eqnarray}}
\def\eea{\end{eqnarray}}

\begin{document}

\title{Exact ground states of a staggered supersymmetric model for lattice fermions}

\author{L. Huijse}
\affiliation{Department of Physics, Harvard University, Cambridge MA 02138, USA}
\author{N. Moran}
\affiliation{Laboratoire Pierre Aigrain, ENS and CNRS, 24 rue Lhomond, 75005 Paris, France}
\affiliation{Department of Mathematical Physics, National University of Ireland, Maynooth, Ireland}
\author{J. Vala}
\affiliation{Department of Mathematical Physics, National University of Ireland, Maynooth, Ireland}
\affiliation{School of Theoretical Physics, Dublin Institute for Advanced Studies, 10 Burlington Road, Dublin 4, Ireland}
\author{K. Schoutens}
\affiliation{Institute for Theoretical Physics, University of Amsterdam, Science Park 904,
P.O.Box 94485, 1090 GL Amsterdam, The Netherlands}

\begin{abstract}
We study a supersymmetric model for strongly interacting lattice fermions in the presence of a staggering parameter. The staggering is introduced as a tunable parameter in the manifestly supersymmetric Hamiltonian. We obtain analytic expressions for the ground states in the limit of small and large staggering for the model on the class of doubly decorated lattices. On this type of lattice there are two ground states, each with a different density. In one limit we find these ground states to be a simple Wigner crystal and a valence bond solid (VBS) state. In the other limit we find two types of quantum liquids. As a special case, we investigate the quantum liquid state on the one dimensional chain in detail. It is characterized by a massless kink that separates two types of order.
\end{abstract}

\maketitle

\section{Introduction}
A central theme in modern day solid state physics is the study of the electronic properties of materials in which the electrons are free to move, but subject to (strong) repulsive interactions. In this regime one typically faces the difficulty that perturbative tools are unreliable. Consequently, even at the level of relatively simple model Hamiltonians, the behavior of such systems is notoriously difficult to analyze.  This challenge motivates the study of a supersymmetric model for strongly interacting, spinless fermions \cite{fendley-2003-90}. In these models a judicious tuning of kinetic and potential terms leads to systems which possess supersymmetry. These models show various interesting features that are closely linked to the supersymmetry. First, supersymmetry provides a rich mathematical structure that allows for a considerable degree of analytic control in the regime where more standard perturbative techniques would fail. Second, it induces a delicate balance between kinetic and potential terms, which results in a strong form of quantum charge frustration. This so-called 'superfrustration', which typically occurs for two (or higher) dimensional lattices, is characterized by an extensive ground state entropy \cite{fendley-2005-95,vanEerten05}. 

The degrees of freedom in the supersymmetric model are itinerant hard-core spin-less fermions. For spin-less fermions double occupancy is forbidden by the Pauli exclusion principle. The hard-core constraint which imposes nearest neighbor exclusion extends this repulsive character. At low enough densities a system of such particles exhibits a metallic phase described by spin-less Fermi liquid theory. On the square lattice the phase at half-filling (one particle every two sites) is trivial; there are two ordered patterns in which the fermions are completely jammed. However, for small hole-doping there is strong evidence that the hard-core spin-less fermions exhibit a stripe phase, where domain walls between the different orders favor being parallel \cite{Henley01,Zhang03,Hizi04}. 

It is clear that the intermediate density regime of this strongly-interacting system is non-perturbative; one has to account for both the kinetic as well as the potential terms. It turns out that the supersymmetric model, which incorporates a next-nearest neighbor repulsion for the itinerant hard-core spin-less fermions, provides a way to study precisely this regime. Exact results for this model on a variety of two (and higher) dimensional lattices have been obtained \cite{fendley-2005-95,Fendley05,vanEerten05,Jonsson06,Jonsson05p,Huijse08b,Huijse10, Huijse10b} (for reviews see \cite{Huijse08a,HuijseT10}). 

A key feature of the model is superfrustration \cite{fendley-2005-95}: a ground state degeneracy that is exponential in the number of lattice sites. One often finds that these ground states occur in a whole region of particle densities. For example, for the supersymmetric model on the square, triangular and honeycomb lattice there exist zero energy ground states for all rational densities $\nu$ within the range [1/5,1/4], [1/7, 1/5] and [1/4, 5/18], respectively \cite{Jonsson05p}. Note that at these densities the model is deep inside the non-perturbative regime. 

A qualitative understanding of superfrustration stems from the 3-rule, which says that particles prefer to be mostly 3 sites apart \cite{fendley-2005-95}. In general it turns out that there is an exponential number of ways to satisfy the 3-rule. For certain lattices, however, the frustration can be dramatically reduced. These lattices have a structure that nicely accomodates the 3-rule. 

On the square lattice the frustration leads to a sub-extensive ground state entropy; the ground state degeneracy is exponential in the linear size of the system. This result is for the square lattice wrapped around the torus, for the cylinder and the plane there is compelling evidence that the system supports quantum critical edge modes \cite{Huijse08b}.

Quantum criticality is also observed for the supersymmetric model in one dimension, where the system turns out to be integrable. The continuum theory is a Luttinger liquid with special values of the Luttinger parameters, such that the theory is supersymmetric \cite{fendley-2003-90,Huijse11}. In recent work \cite{Fendley10a,Fendley10b}, it was shown that the one dimensional model also enjoys special features, such as scale-free properties, which are again directly related to supersymmetry.

In this paper we study the supersymmetric model extended with a tunable parameter, which for reasons that will become clear later, we call the staggering parameter \cite{fendley-2003-36,fendley-2005-95,Fendley10a,Fendley10b}. We show that for the lattices with reduced frustration we can obtain analytic expressions for the zero energy states in the limits where this parameter is very large and very small.

Interestingly, we find a whole class of graphs where, in the limit of a very small staggering parameter, the ground state is the equal weight superposition of all states in the low energy sector. Such liquid states typically arise in the context of quantum dimer models, where this is a consequence of the Perron-Frobenius theorem. Famous models with liquid ground states are Kitaev's toric code \cite{Kitaev03} and Levin and Wen's string-nets \cite{Levin05}, which exhibit topological order, and the Rokhsar-Kivelson (RK) model \cite{Rokhsar88}, which exhibits criticality. The class of models we discuss in this paper, however, seem to be of a different type. We find that for a particular choice of lattice, our effective Hamiltonian precisely maps to the constrained quantum-clock model that arises in the context of bosons in a tilted optical lattice \cite{Pielawa11}. For this model it has been shown that the ground state has no long range order, which suggests that it is in a gapped phase. In our model this quantum liquid state has one particle per plaquette. We find that there is another quantum liquid ground state at a higher density, which has one particle per link. The two quantum liquid states arise from a projection of a product state of single particle states with one particle per vertex plaquette and one particle per link respectively. The procedure is reminiscent of the construction of the Affleck-Kennedy-Lieb-Tasaki (AKLT) state \cite{aklt87,aklt88}. Besides the quantum liquids, we will find that the model also supports other gapped phases. One being a charge ordered crystalline phase, the other a valence bond solid (VBS) that does not break any lattice symmetry. In contrast to the quantum liquids, the ground states in the latter two phases are simple product states.

As a special case of the class of models with liquid ground states we discuss the staggered supersymmetric model on the one dimensional chain with open boundary conditions. It is well known \cite{fendley-2003-90,Huijse11} that the supersymmetric model on the chain without staggering is quantum critical. In \cite{Fendley10a} it was shown that the staggering parameter corresponds to a relevant operator in the continuum theory. In this paper we show that in one limit of the staggering parameter, the ground state of the chain with open boundary conditions is a liquid state. This liquid state contains a delocalized massless kink. We will see that, as a consequence, the spectrum in that limit has a zero dimensional massless character.

In the last part of the paper, we speculate that for lattices that do not have reduced frustration an analysis in the same spirit as the one presented here may lead to the discovery of other exotic phases. There are indications for RK type phases and the existence of one dimensional critical modes in two dimensional lattices.

In the remainder of this introduction we first discuss the basics of supersymmetry for lattice fermions. We then define the model and discuss the consequences of the staggering parameter. Finally, we introduce cohomology. This mathematical technique has proven extremely powerful in the analysis of the supersymmetric model and will also be essential for the analyses in this work.

\subsection{Supersymmetry for lattice fermions}
Supersymmetry is a symmetry between fermionic and bosonic
degrees of freedom (see \cite{Bagger03} for a general reference).
It plays an important role in theoretical high energy physics,
where various theories that go beyond the standard model require
supersymmetry for a consistent formulation. In these theories all
the known elementary particles are accompanied by yet to be
discovered superpartners.

In the lattice model discussed here the physical particles are
spinless lattice fermions and the supersymmetry relates 
fermionic and bosonic many particle states with an odd and
even number of the lattice fermions respectively. In the $\mathcal{N}=2$ supersymmetric theories that we consider, a central role
is played by the operators $Q$ and $Q^{\dag}$, called
supercharges, which have the following properties \cite{Witten82}
\begin{itemize}
\item $Q$ adds one fermion to the system and $Q^{\dag}$ takes out
one fermion from the system. 
\item The supercharges are fermionic
operators and thus nilpotent: $Q^2 =(Q^{\dag})^2 =0$. 
\item The
hamiltonian is the anti-commutator of the supercharges, $H =
\{Q,Q^{\dag}\}$, and as a consequence it commutes with the
supercharges and conserves the number of fermions in the system.
\end{itemize}

Imposing this structure has some immediate consequences; supersymmetric theories are characterized by a positive 
definite energy spectrum and a twofold degeneracy of
each non-zero energy level. The two states with the same energy are called 
superpartners and are related by the supercharge. 

From its definition it follows directly that $H$ is positive definite:
\begin{eqnarray*}
\bra{\psi}H\ket{\psi}
  &=& \bra{\psi}(Q^{\dag}Q+QQ^{\dag})\ket{\psi}
\nonumber \\[2mm]
  &=&|Q\ket{\psi}|^2+|Q^{\dag}\ket{\psi}|^2 \geq 0 \ .
\end{eqnarray*}
Furthermore, both $Q$ and $Q^{\dag}$ commute with the Hamiltonian, 
which gives rise to the twofold degeneracy in the energy spectrum. 
In other words, all eigenstates with an energy $E_s>0$ form doublet
representations of the supersymmetry algebra. A doublet consists of two 
states $\ket{s}, Q \ket{s}$, such that $Q^{\dag}\ket{s}=0$. Finally, 
all states with zero energy must be singlets: 
$Q \ket{g}=Q^{\dag}\ket{g}=0$ and conversely, all singlets must be zero 
energy states \cite{Witten82}. In addition to supersymmetry our models also 
have a fermion-number symmetry generated by the operator $F$ with
\begin{eqnarray}\nonumber
[F,Q^{\dag}]=-Q^{\dag} \quad \textrm{and} \quad [F,Q]=Q.
\end{eqnarray}
Consequently, $F$ commutes with the Hamiltonian. From the commutators of $F$ with the supercharges, one finds that $Q$ 
and $Q^{\dag}$ change the number of fermions by plus or minus one unit, so that the 
supercharges map bosonic states to fermionic states and vice versa.

\subsection{The model}
We now make things concrete and define supersymmetric models for spin-less
fermions on a lattice or graph with $L$ sites in any dimension, following
\cite{fendley-2003-90}. The operator 
that creates a fermion on site $i$ is written as $c_i^{\dag}$ with
$\{c_i^{\dag},c_j\}=\delta_{ij}$. To obtain a non-trivial Hamiltonian, we dress the fermion 
with a projection operator: $P_{\langle i \rangle}=\prod_{j \textrm{ next to } i} (1-c_j^{\dag}c_j)$, which requires 
all sites adjacent to site $i$ to be empty. We now define the supercharges as
\bea
Q=\sum \lambda^{*}_i c_i^{\dag} P_{\langle i \rangle}, \quad Q^{\dag}=\sum \lambda_i c_i P_{\langle i \rangle}.
\eea
One can check that the supercharges are nilpotent for any complex numbers $\lambda_i$.
It follows that the Hamiltonian of these hard-core fermions reads
\begin{eqnarray}\label{Hsusygen}\nonumber
H=\{Q^{\dag},Q\}= 
\sum_i \sum_{j\textrm{ next to }i} \lambda_i^{*} \lambda_j P_{\langle i \rangle} c_i^{\dag} c_j P_{\langle j \rangle} 
       + \sum_i |\lambda_i|^2 P_{\langle i \rangle}.
\end{eqnarray}
The second term contains a next-nearest neighbor repulsion, a chemical 
potential and a constant. The details of the latter terms will depend on the lattice we choose.
The first term is just a nearest neighbor hopping term for hard-core fermions. The projection operators in this term ensure that the nearest neighbor exclusion is obeyed by the hard-core fermions. Another way to impose this condition would be to include a nearest neighbor repulsion of strength $V$ and send $V$ to infinity. 

Physical systems which realize this type of interactions are ultra cold gasses of Rydberg atoms. The Van der Waals interaction between Rydberg atoms falls off as $1/r^{6}$, where $r$ is the distance between two Ryberg atoms. It follows that the nearest neighbor repulsion is almost two orders of magnitude larger than the next-nearest neighbor repulsion. This results in what is called a Rydberg excitation blockade; atoms within a certain distance of a Rydberg atom, cannot be excited to the Rydberg state. It follows that Rydberg atoms can be accurately modelled by hard-core fermions.

In this paper we focus on lattices $S$ which naturally break up into two sublattices $S_1$ and $S_2=S \backslash S_1$. On these lattices we choose the following staggering:
\beq\label{eq:stagparam}
\lambda_i=\left\{ \begin{array}{ll} 1 & \textrm{for $i \in S_1$}\\
y & \textrm{for $i \in S_2$} \end{array} \right.
\eeq
It follows that the hopping amplitudes between two $S_1$ sites, an $S_1$ and an $S_2$ site and two $S_2$ sites are 1, $y$ and $y^2$ respectively. Furthermore, we find that the chemical potential on the sublattice $S_1$ is independent of $y$, whereas the chemical potential on $S_2$ is proportional to $y^2$. The consequences of the staggering on the interaction terms is a bit more subtle and will depend on the details of the lattice. 

\subsection{Cohomology}
For the supersymmetric models, cohomology has proven to be a very powerful tool to extract information about the zero energy ground state(s) of the models (see for example \cite{fendley-2003-90,fendley-2005-95,Jonsson05p,Huijse08b,Huijse10, HuijseT10} and references therein). The key ingredient is the fact that ground states are singlets; they are annihilated both by $Q$ and $Q^{\dag}$. This means that a ground state $\ket{g}$ is in the kernel of $Q$:  $Q\ket{g}=0$ and not in the image of $Q$, because if we could
write $\ket{g}=Q\ket{f}$, then $(\ket{f}, \ket{g})$, would be a doublet. So the ground states span a subspace $\mathcal{H}_Q$ of the Hilbert space $\mathbf{H}$ of states, such that $\mathcal{H}_Q=\ker Q/\im Q$. This is precisely the definition of the cohomology of $Q$. So the ground states of a supersymmetric theory are in one-to-one correspondence with the cohomology of $Q$. It follows that the solution of the cohomology problem gives the number of zero energy states for each particle number sector. 

We compute the cohomology using the `tic-tac-toe' lemma of \cite{BottTu82}. This says that under certain conditions, the cohomology
$\mathcal{H}_Q$ for $Q = Q_1 + Q_2$ is the same as the cohomology of $Q_1$ acting on the cohomology of $Q_2$. In an equation, $\mathcal{H}_Q = \mathcal{H}_{Q_1}(\mathcal{H}_{Q_2} ) \equiv \mathcal{H}_{12}$, where $Q_1$ and $Q_2$ act on different sublattices $S_1$ and $S_2$. We find $\mathcal{H}_{12}$ by first fixing the configuration on all sites of the sublattice $S_1$, and computing the cohomology $\mathcal{H}_{Q_2}$. Then one computes the cohomology of $Q_1$, acting not on the full space of states, but only on the classes
in $\mathcal{H}_{Q_2}$. A sufficient condition for the lemma to hold is that all non-trivial elements of $\mathcal{H}_{12}$ have the same $f_2$ (the fermion-number on $S_2$).

Although ground states are in one-to-one correspondence with cohomology elements, the two are not equal unless the cohomology element happens to be a harmonic representation of the cohomology. Harmonic representations are elements of the cohomology of both $Q$ and $Q^{\dag}$. So they are annihilated by both supercharges, which is precisely the property of a zero energy state. It follows that, although the solution of the cohomology problem gives the number of ground states, it typically does not give the ground states themselves. 
In this paper, however, we show that for the lattices with reduced frustration the cohomology computation gives the exact ground states in the limits where the staggering is very large and very small. This property was also successfully exploited in \cite{Fendley10a,Fendley10b}. The fact that cohomology elements can provide more information about the ground states than just their degeneracy and particle density, was also observed in \cite{Huijse08b,Huijse10b,HuijseT10}.

\section{Cohomology for staggered lattices}\label{sec:cohom1}
In this paper we investigate the properties of the supersymmetric model in the limits of very small and very large staggering. It follows that we can treat the problem perturbatively in either $y$ or $1/y$. As a first step we compute the ground states to zeroth order in the two limits. When we find that the number of zeroth order ground states is larger than the number of ground states at finite $y$, the second step is to do degenerate perturbation theory. We find that the degeneracy is lifted at second order in perturbation theory. In both steps cohomology plays an essential role. 

First of all, we will find the zeroth order ground states by solving the cohomology problem in the two limits. Remember that we introduced the staggering parameter $y$, such that the supercharge takes the form $Q = Q_1 + y Q_2$ and the Hamiltonian reads $H=\{ Q, Q^{\dag}\}=H_0+y H_1 + y^2 H_2$. It follows that for $y=0$ we have $Q=Q_1$ and for $y \to \infty$ we have $Q \approx y Q_2$. For the lattices we consider, we can easily compute both $\mathcal{H}_{Q_1}$ and $\mathcal{H}_{Q_2}$ and we will find that the elements of $\mathcal{H}_{Q_1}$ and $\mathcal{H}_{Q_2}$ respectively are precisely the zeroth order ground states in the small and large staggering limit. 

Second of all, cohomology tells us that if the number of ground states in either of the two limits is larger than the number of ground states for finite $y$, this degeneracy is lifted at second order in perturbation theory. Typically, we will find that $\mathcal{H}_{Q_1}$ contains many more elements than $\mathcal{H}_{Q}$, that is, $D_1>D$, with $D_1 = \dim \mathcal{H}_{Q_1}$ and $D=\dim \mathcal{H}_Q$. It follows that for $y$ strictly zero there are $D_1$ zero energy states. We will now prove that $D_1-D$ of these states will acquire an energy of order $y^2$ when $y$ is no longer strictly zero, but still small. An important condition for this to be true is that we have that $\mathcal{H}_Q = \mathcal{H}_{Q_2}(\mathcal{H}_{Q_1} )$, that is, the 'tic-tac-toe' lemma should hold.

Let $\ket{\chi}$ be an eigenstate of the Hamiltonian, $H$, with energy $E(y)>0$. Furthermore, we assume that $H_0 \ket{\chi_0}=0$, where we defined $\ket{\chi_0}\equiv \lim_{y \to 0} \ket{\chi}$. It follows that $\ket{\chi_0} \in \mathcal{H}_{Q_1}$ and $\ket{\chi_0} \notin \mathcal{H}_Q$. We will assume that $\mathcal{H}_Q = \mathcal{H}_{Q_2} (\mathcal{H}_{Q_1})$. Since $\ket{\chi_0} \notin \mathcal{H}_{Q_2} (\mathcal{H}_{Q_1})$ we may write $Q_2 \ket{\chi_0}= \ket{\phi}$ and $Q_2^{\dag} \ket{\chi_0}=\ket{\phi'}$, where $\ket{\phi}$ and $\ket{\phi'}$ are not both zero. 

It then follows that the energy of $\ket{\chi_0}$ is
\bea
E (y) &\approx& \bra{\chi_0} H \ket{\chi_0} \nn
&=& \bra{\chi_0} H_0 \ket{\chi_0}+ y \bra{\chi_0} H_1 \ket{\chi_0} + y^2 \bra{\chi_0} H_2 \ket{\chi_0} \nn
&=& y \bra{\chi_0} (Q_1 Q_2^{\dag} + Q_1^{\dag} Q_2 ) \ket{\chi_0} + y^2  \bra{\chi_0} (Q_2 Q_2^{\dag} + Q_2^{\dag} Q_2 ) \ket{\chi_0} \nn
&=& y  (Q_1^{\dag} \ket{\chi_0})^{\dag} \ket{\phi'}+ y  (Q_1 \ket{\chi_0})^{\dag} \ket{\phi} + y^2  (Q_2^{\dag} \ket{\chi_0})^{\dag} \ket{\phi'}+ y^2  (Q_2 \ket{\chi_0})^{\dag} \ket{\phi}\nn
&=& y^2 ( \braket{\phi'}{\phi'} + \braket{\phi}{\phi} ) \nn
&=& N y^2, \nonumber
\eea
where $N>0$. This proves our statement above.

In this paper we will exploit the fact that for any finite $y$ the degeneracy at $y=0$ will be lifted at second order in perturbation theory. As we will see the effective Hamiltonian that couples the cohomology elements of $\mathcal{H}_{Q_1}$ will take on some simple form and the zero energy states can be found exactly. 

\section{Exact ground states for staggered chain}\label{sec:chaings}
In this section we consider the staggered one dimensional chain with open boundary conditions and length $L=3j$. The methods explained for this special case will carry over to a variety of two dimension lattices in the next section. We discuss the cohomology problem in general and in the two limits of large and small staggering. We find that for all finite values of the staggering parameter the ground state is unique and has particle number $f=j$. For $y$ strictly zero we find that there are $2j+1$ zero energy states. For finite staggering this degeneracy is lifted. The effective Hamiltonian that couples these states is obtained in second order degenerate perturbation theory. The ground state of the effective Hamiltonian is an equal superposition liquid state.

We will use the following conventions. For a chain of length $L=3j$ we number the sites as $i=0,\dots, 3j-1$ and for the staggering we choose:
\beq
\lambda_i=\left\{ \begin{array}{ll} 1 & \textrm{for $i=1$ mod $3$}\\
y & \textrm{otherwise.} \end{array} \right.
\eeq

\subsection{Cohomology for staggered chain}\label{subsec:chaincohom}
To compute the cohomology for the chain, one uses the fact that the cohomology of a single site that can be both empty and occupied, is trivial. This is equivalent to the statement that the single site chain has no zero energy states, indeed the empty and occupied state form a doublet of energy one. For the chain of length $L=3j$ with open boundary conditions, we define the sublattice $S_1$ as all sites $i=1 \mod 3$ (remember that we label the sites as $0 \leq i \leq L-1$). Sublattice $S_2$ consists of the remaining sites. Solving $\mathcal{H}_{Q_2}$ is easy: it is trivial unless the sites 0 and $L-1$ are blocked by occupying the sites 1 and $L-2$, but now $\mathcal{H}_{Q_2}$ is still trivial unless the sites 3 and $L-4$ are blocked, etc. Continuing this reasoning we find there is only one non-trivial element, which has all $S_1$ sites occupied. Since there is only one element in $\mathcal{H}_{Q_2}$, finding $\mathcal{H}_{12}$ is trivial: $\mathcal{H}_{12}=\mathcal{H}_{Q_2}$. For the same reason the 'tic-tac-toe' lemma holds and we find that $\mathcal{H}_{Q}$ consists of one non-trivial element. It follows that there is one zero energy state with $j$ fermions.

The above argument holds for all values of the staggering parameter, except when $y=0$. In that case $Q_2$ drops out alltogether so it does not make sense to compute $\mathcal{H}_{Q_2}$. Instead we have $\mathcal{H}_{Q}=\mathcal{H}_{Q_1}$, which we can compute directly. Since $S_1$ is a collection of isolated sites, the cohomology of $Q_1$ vanishes, unless all $S_1$ sites are blocked. One easily verifies that there are $j+1$ configurations with $j$ fermions and $j$ configurations with $j+1$ fermions for which all $S_1$ sites are blocked. If we denote the sites $3i+1$ by a square and the other sites by a circle and color occupied sites black, we can depict these configurations as
\bea\label{eq:cohomelemchain}
\begin{array}{ccc}
 \bullet \square \circ  \bullet \square \circ \dots  \bullet \square \circ \bullet \square \circ & \quad & \bullet \square \circ  \bullet \square \circ \dots  \bullet \square \circ \bullet \square \bullet \\
 \bullet \square \circ  \bullet \square \circ \dots  \bullet \square \circ  \circ \square \bullet & \quad & \bullet \square \circ  \bullet \square \circ \dots  \bullet \square \bullet  \circ \square \bullet \\
 \vdots & \quad & \vdots \\
\bullet \square \circ  \circ \square \bullet  \dots  \circ \square \bullet \circ \square \bullet & \quad & \bullet \square  \bullet \circ \square \bullet  \dots  \circ \square \bullet \circ \square \bullet \\
\circ \square \bullet   \circ \square \bullet  \dots  \circ \square \bullet \circ \square \bullet & \quad & 
\end{array}
\eea

It follows that for $y$ strictly zero, there are $2j+1$ zero energy states. In fact, in this limit the zero energy states are precisely the states depicted above. For $y\to \infty$ the zero energy state is the state with all $S_1$ sites occupied:
\beq\label{eqn:gs_obc_large}
\ket{\psi_{\textrm{GS}}(y \to \infty)} =  \prod_{k=0}^{j-1}c^{\dagger}_{3k+1} \ket{\emptyset}.
\eeq

Finally,  we note that the Witten index, which is the trace over the ground state manifold of $(-1)^F$, remains unchanged for all values of $y$: $W=(-1)^j$ for $y>0$ and $W=(j+1)(-1)^j+j(-1)^{j+1}=(-1)^j$ for $y=0$.

\subsection{Effective Hamiltonian for finite staggering}\label{sec:deglift} 
In this section we discuss degenerate perturbation theory in the limit of small staggering parameter $y$. Since the supercharges are linear in $y$ the Hamiltonian takes the form $H=H_0+y H_1 +y^2 H_2$. Here $H_0$ is purely diagonal, $H_1$ is purely off-diagonal, it is the hopping between staggered and non-staggered sites, and finally, $H_2$ consist of a diagonal and an off-diagonal part, where the off-diagonal part represents hopping between the sites $3i$ and $3i-1$ for $1 \leq i \leq L/3-1$.

We know that at finite staggering there is a unique ground state with zero energy and $f=j$. So let us write for this zero energy state
\bea
\ket{\Psi}=\ket{\Psi_0}+y\ket{\Psi_1}+y^2 \ket{\Psi_2} + \dots
\eea
 where $\ket{\Psi_0}= \sum_{k=0}^{j} a_k \ket{\psi_{k}}$. The configurations $\ket{\psi_k}$ are the cohomology elements of $\mathcal{H}_{Q_1}$ with $j$ particles as shown in (\ref{eq:cohomelemchain}). They can be defined as:
\bea\label{eq:psik}
\ket{\psi_{i}} = \prod_{k=0}^{i-1} c_{3k}^{\dagger} \prod_{k=i}^{j-1}c_{3k+2}^{\dagger} \ket{\emptyset},
\eea
for $0\leq i\leq j$. Similarly, we write for the cohomology elements of $\mathcal{H}_{Q_1}$ with $j+1$ particles:
\bea\label{eq:phik}
\ket{\phi_{i}} = \prod_{k=0}^{i} c_{3k}^{\dagger} \prod_{k=i}^{j-1}c_{3k+2}^{\dagger} \ket{\emptyset},
\eea
with $0\leq i\leq j-1$.

 To determine the coefficients $a_k$ we have to go to second order perturbation theory. To first order we have
\bea\label{eq:1stord}
H_0 \ket{\Psi_1} = (E_1 - H_1) \ket{\Psi_0}.
\eea
Taking the inner product on both sides with $ \ket{\psi_{l}}$, we find
\bea
0= \bra{\psi_l}(E_1 - H_1)\sum_{k=0}^{j} a_k \ket{\psi_{k}} = E_1 a_l .
\eea
Here we used $\bra{\psi_l}H_1\ket{\psi_k}=0$, since $H_1$ is purely off-diagonal and only has non-zero elements between configurations which differ in their number of particles on the sites $3i+1$ by one. Clearly, we cannot have that all $a_l$ are zero, so we find $E_1=0$ and the degeneracy is not lifted. 

To find the wave function to first order we write $\ket{\Psi_1}= \sum_m a^1_m \ket{\chi_m}$, where $H_0 \ket{\chi_m} = E_m  \ket{\chi_m}$ with $E_m>0$. We then take the inner product on both sides in (\ref{eq:1stord}) with $\ket{\chi_l}$. This way we find
\bea\label{eq:psi1coeff}
a^1_l= - \sum_{k=0}^{j} a_k \frac{\bra{\chi_l} H_1 \ket{\psi_k}}{E_l}.
\eea

Now to second order we have
\bea\label{eq:2ndord}
H_0 \ket{\Psi_2} = (E_1 - H_1) \ket{\Psi_1} + (E_2-H_2)\ket{\Psi_0}.
\eea
Again taking the inner product with $ \ket{\psi_{l}}$ and using $E_1=0$, we find
\bea
0&=& \bra{\psi_l}(E_1 -H_1) \ket{\Psi_1} +   \bra{\psi_l}(E_2 -H_2) \ket{\Psi_0}\nonumber\\
&=& \sum_{k=0}^{j} \left[  \sum_m a_k \frac{ \bra{\psi_l} H_1 \ket{\chi_m}\bra{\chi_m} H_1 \ket{\psi_k}}{E_m} -  a_k  \bra{\psi_l}  H_2  \ket{\psi_{k}} \right] + a_l E_2
\eea
These homogeneous algebraic equations can only be solved for the coefficients $a_l$ if the determinant of their coefficients vanishes. The resulting secular equation reads
\bea
|W_{lk} - \delta_{lk }E_2| =0 ,
\eea
where
\bea
W_{lk} = - \sum_m \frac{ \bra{\psi_l} H_1 \ket{\chi_m}\bra{\chi_m} H_1 \ket{\psi_k}}{E_m} + \bra{\psi_l}  H_2  \ket{\psi_{k}}.
\eea
The matrix $W$ is interpreted as an effective Hamiltonian that couples the states $\ket{\psi_k}$. It can easily be shown that $W$ takes on the simple form
\bea\label{eq:Heffchain}
W&=& \left( \begin{array}{ccccc}
-j & -1 & &  &  \\
-1 & & &\O &  \\
& & \ddots & &  \\
 &\O  & &  & -1 \\
& & & -1 & -j \end{array} \right) + \left( \begin{array}{ccccc}
j+1 &  & &  &  \\
 & j+2 & &\O &  \\
& & \ddots & &  \\
 &\O  & & j+2 &  \\
& & &  & j+1 \end{array} \right) \nonumber\\
&=& \left( \begin{array}{ccccc}
1 & -1 & &  &  \\
-1 & 2 & &\O &  \\
& & \ddots & &  \\
 &\O  & & 2 & -1 \\
& & & -1 & 1 \end{array} \right) 
\eea
Note that the rows of this effective Hamiltonian all sum to zero. Consequently, the zero energy ground state is an equal superposition of all the $\ket{\psi_k}$:
\beq\label{eq:gs0th}
\ket{\Psi_0}= \sum_{k=0}^{j} \ket{\psi_{k}}.
\eeq
Remember that we showed in section \ref{sec:cohom1}, that at second order in perturbation theory $D_1-D$ states will acquire an energy of order $y^2$. In the sector with $j$ particles we have $D_1=\dim \mathcal{H}_{Q_1}=j+1$ and $D=\dim \mathcal{H}_Q =1$. From this it follows that the $(j+1)\times(j+1)$ matrix $W$ has a unique zero energy state. Another way to see that $\ket{\Psi_0}$ is the unique ground state is via the Perron-Frobenius theorem. Note that all off-diagonal elements of $W$ are negative and that repeated application of this effective Hamiltonian connects all configurations. From these two properties it follows by the Perron-Frobenius theorem that the ground state is unique and nodeless. For this very simple matrix, using the Perron-Frobenius theorem is somewhat pedantic. In fact, as we shall discuss in section \ref{sec:kink}, we can obtain all eigenvalues and eigenstates of this system. However, for the two dimensional lattices that we discuss in the next paragraph we will see that many of the effective Hamiltonians share this property.

We have found that the degeneracy of the states with $j$ particles is lifted in second order perturbation theory. Due to supersymmetry it is now directly clear that the degeneracy of the states $\ket{\phi_i}$ with $f=j+1$ is also lifted. This is because supersymmetry dictates that the $j$ states with energies $y^2 E_2>0$ have superpartners with equal energy and $f=j\pm1$. Clearly, these superpartners must be linear superpositions of the states $\ket{\phi_i}$, because all other states have energies of order one. Since there are $j$ such states it follows that there are no zero energy states at $f=j+1$. We conclude that there is a unique zero energy state, which has $f=j$ particles.

\section{Exact ground states for staggered decorated lattices}
We now turn to two (or higher) dimensional lattices. In this section we apply the same approach that we used in the previous section for the chain to a whole class of graphs (a graph is just a lattice, but the unit cell may contain all sites). The graphs we consider are defined as follows. We start with an original graph $\Lambda$ and construct the graph $\Lambda_n$ from the original graph by adding $n-1$ additional vertices on every link. In the following the only restriction on the graph $\Lambda$ is that it does not contain disconnected subgraphs. It is clear that these decorated graphs, $\Lambda_n$, naturally break up into two sublattices. We have seen that the  'tic-tac-toe' lemma is then especially powerful and in the following we will see how this property can be exploited.

We will first discuss the cohomology problem for the decorated graphs in general \cite{fendley-2005-95,Csorba09,HuijseT10}. We will then restrict our attention to the doubly decorated graphs, $\Lambda_3$. For these graphs we will show that upon staggering we can find quantum liquid, crystalline and VBS type states.

\subsection{Cohomology for decorated graphs}
The cohomology problem for the graphs of type $\Lambda_{3m}$ can be solved for general $m$. For graphs of type $\Lambda_{3m\pm1}$ there is no general solution to the cohomology problem, however, one can relate it to the cohomology problem of the original lattice $\Lambda$.  These results were obtained on a homotopy level using Alexander dualities in \cite{Csorba09}, but can also be derived using the 'tic-tac-toe' lemma \cite{HuijseT10}.

\begin{figure}
\centering
\includegraphics[width=0.4\textwidth]{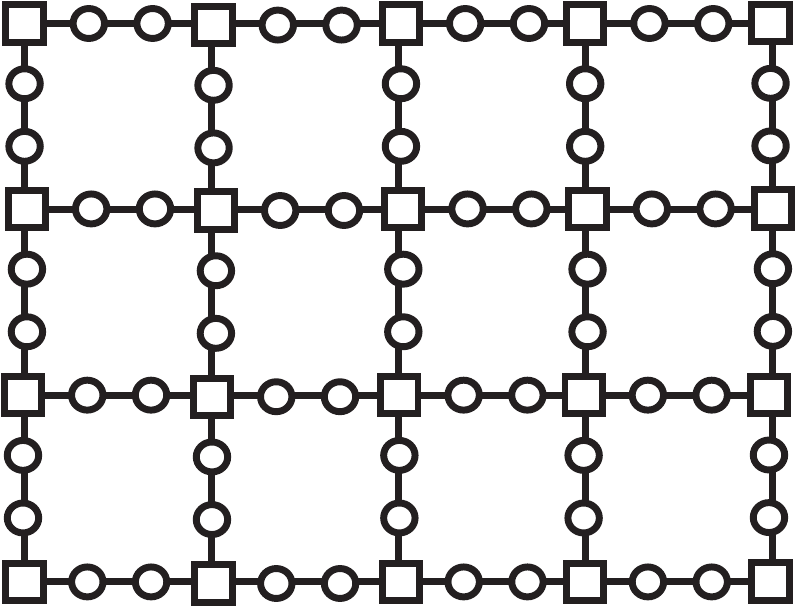}
\caption{Decorated square lattice. The sublattice $S_1$ is the original square lattice, the sites are depicted as squares. The sublattice $S_2$ are the added sites, indicated by circles, on the links of the original lattice. For doubly periodic boundary conditions the right (bottom)
sites are identified with the left (top) sites.}
\label{fig:Sq3}
\end{figure}

Let us resume the cohomology problem of the graph $\Lambda_3$, which was discussed in
\cite{fendley-2005-95}. In the following the original graph $\Lambda$ can have open or closed boundary conditions, but we assume that no changes are made to the boundary conditions once the additional sites are added to the links of $\Lambda$. We postpone the discussion on how the results are modified when the boundary conditions are changed by cutting links in the $\Lambda_3$ graph to the end of this section. Consider the original graph $\Lambda$ as the
subgraph $S_1$ and the additional sites as the subgraph $S_2$. The graph $S_2$ is a collection of
two site chains. Remember that an isolated site that can be both empty and occupied has a trivial cohomology. It follows that the non-trivial elements in $\mathcal{H}_{Q_2}$ have the $S_1$ sites neighboring a
two site chain on $S_2$ either both empty or both occupied. It follows that
there are only two non-trivial elements in $\mathcal{H}_{Q_2}$, one with $S_1$
completely empty and one with $S_1$ completely filled. If we leave
$S_1$ completely empty, we obtain one non-trivial element of $\mathcal{H}_{Q_2}$
in the sector with $L_{\Lambda}$ fermions, where $L_{\Lambda}$ denotes the number of
links in the original graph $\Lambda$. The element with $S_1$
completely filled, clearly has fermion number $N_{\Lambda}$, where
$N_{\Lambda}$ denotes the number of vertices in the original graph
$\Lambda$. It now quickly follows that these two elements are also
in $\mathcal{H}_{12}$, since within $\mathcal{H}_{Q_2}$ both states are in the kernel of $Q_1$ and not in the image of $Q_1$. Now remember that for the 'tic-tac-toe' lemma to hold a sufficient condition was that all elements in $\mathcal{H}_{12}$ have the same number of fermions on sublattice $S_2$. Here this condition is clearly not met. One can easily show, however, that the lemma also holds when the elements of $\mathcal{H}_{12}$ do not differ in their total fermion number by one. We thus conclude that $\mathcal{H}_{12}=\mathcal{H}_Q$ provided that $N_{\Lambda} \neq L_{\Lambda} \pm 1$.

As an example consider the square lattice with doubly periodic boundary conditions as the original graph $\Lambda$. The lattice $\Lambda_3$ is shown in figure \ref{fig:Sq3}. We find that $L_{\Lambda}=2 N_{\Lambda}$ and the total number of sites in $\Lambda_3$ is $N=2L_{\Lambda}+ N_{\Lambda}$. Consequently, this lattice has one ground state at $1/5$ filling and one at $2/5$ filling.

Before we continue, a few remarks about the possible boundary conditions are in place. If we have periodic boundary conditions on the graph $\Lambda_3$ there is no ambiguity, however, there are a few inequivalent possibilities for open boundary conditions. On the one hand, we can consider the case that the original graph $\Lambda$ has open boundary conditions, constructing the $\Lambda_3$ graph is then unambiguous. On the other hand, we can have periodic boundary conditions on the original graph, construct the $\Lambda_3$ graph and only then cut various links to impose open boundary conditions. In the $\Lambda_3$ graph, however, there are inequivalent links; there are links that connect two $S_2$ sites and links that connect an $S_1$ site to an $S_2$ site. In principle we want to allow cutting both types of links, but one has to be careful, because it has consequences for the solution to the cohomology problem. When open boundary conditions are imposed by cutting links that connect $S_2$ sites, the cohomology in the sector with $L_{\Lambda}$ fermions vanishes. This is because the $S_2$ lattice now contains disconnected sites, so the configuration with all $S_1$ sites empty is no longer an element of $\mathcal{H}_{Q_2}$. Note that the chain with open boundary conditions and length a multiple of three is an example of a $\Lambda_3$ graph with this type of boundary conditions. In section \ref{subsec:chaincohom} we indeed found that it has only one zero energy state. Conversely, when open boundary conditions are imposed by cutting the other type of links, the configuration with all $S_1$ sites occupied is no longer an element of $\mathcal{H}_{Q_2}$. Consequently, for that type of boundary conditions there is no ground state with $N_{\Lambda}$ particles.

\begin{figure}
\centering
\includegraphics[width=0.6\textwidth]{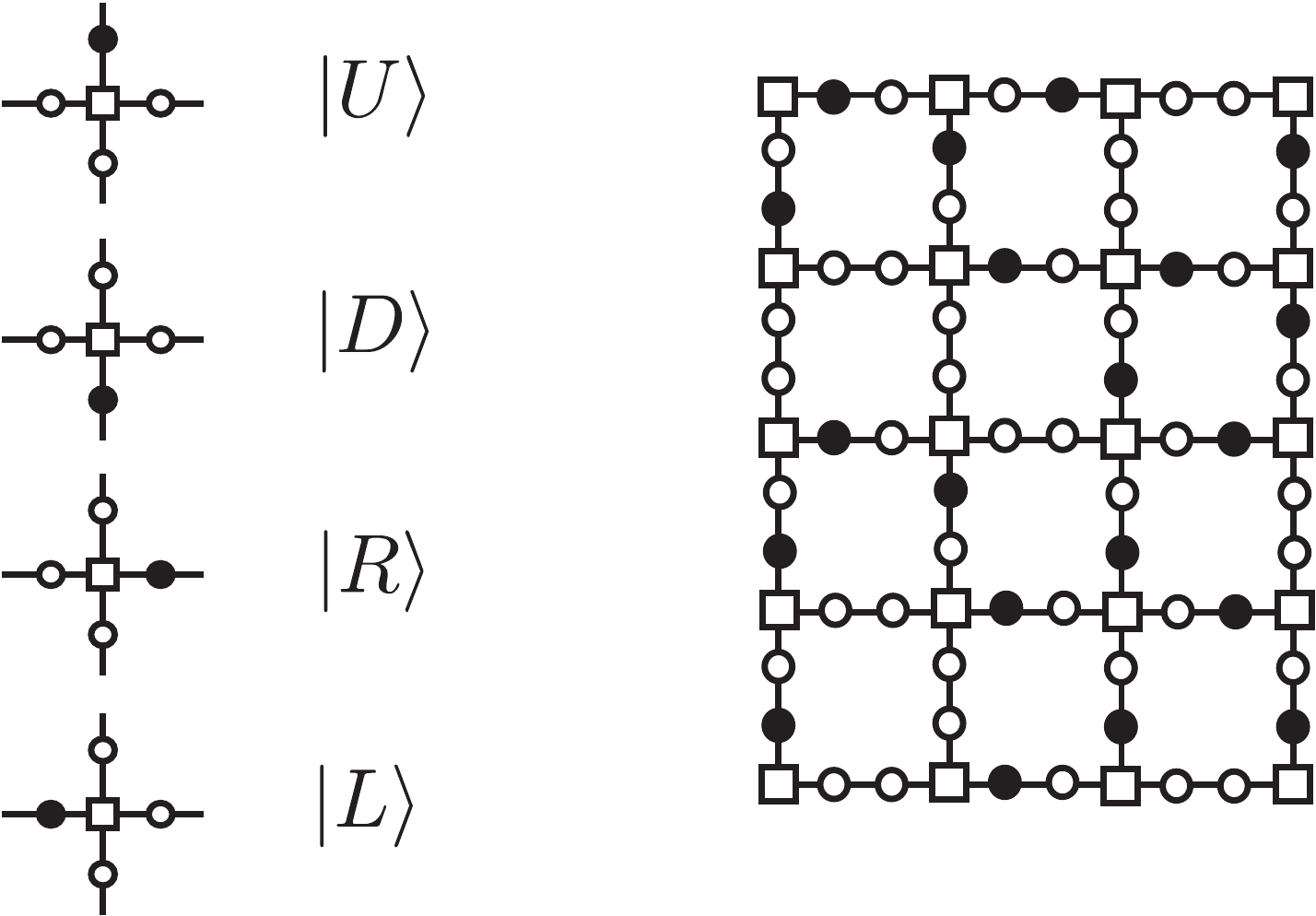}
\caption{Mapping of elements of the cohomology of $Q_1$ at $F=N_{\Lambda}$ to projected clock model states. On the left we show the four possible states of each vertex. On the right we show, as an example, a cohomology element of $\mathcal{H}_{Q_1}$ with $F=N_{\Lambda}$ for the doubly decorated square lattice with open boundary conditions.}
\label{fig:clockmap}
\end{figure} 

\subsection{Ground states to zeroth order}
Let us now introduce the staggering parameter $y$ in these models as defined in (\ref{eq:stagparam}). First consider the limit of infinite staggering. To zeroth order in $1/y$ the Hamiltonian reads $H/y^2 \approx H_2 = \{ Q^{\dag}_2 , Q_2 \} $. The two elements of $\mathcal{H}_{Q_2}$, one with $S_1$ completely empty and one with $S_1$ completely filled, are precisely the ground states of this Hamiltonian. In the state with $S_1$ completely filled the particles force all $S_2$ sites to be empty. Consequently, this state is annihilated by both $Q^{\dag}_2$ and $Q_2$. The state is a simple product state, so the correlation length is finite. We call this charge ordered phase a Wigner crystal. 

Now consider the other state which has $S_1$ completely empty. The $S_2$ sublattice is a collection of two-site chains. The ground state of a two-site chain contains one particle that resonates between the two sites: $\ket{\psi_0}=(c_1^{\dag}-c^{\dag}_2)/\sqrt{2}\ket{0}$. It follows that in the state with $S_1$ completely empty there is one particle on each two-site chain of $S_2$ sites that resonates between the two sites. So also this state is a product state. For obvious reasons we call this gapped phase a VBS.

It is interesting to note that although both states are quite simple, they would have been hard to find without the cohomology solution. Clearly, the charge ordered state is the expected ground state when the system is perturbed with a large conventional staggering that simply favors the $S_1$ sites to be occupied. We thus expect this state to dominate a significant part of the phase diagram also when the parameters are such that the model is no longer supersymmetric. The VBS state, on the other hand, is the energetically favored state when a large conventional staggering favors the $S_2$ sites to be occupied. It is remarkable that for the staggering that preserves supersymmetry the two states coexist. 

We now turn to the other limit. For $y=0$ we have $Q=Q_1$ and we should solve the cohomology problem by looking at $\mathcal{H}_{Q_1}$ directly. Since $S_1$ is a collection of disconnected sites, the non-trivial cohomology elements of $\mathcal{H}_{Q_1}$ are all configurations which have at least one neighboring site of each $S_1$ site occupied. There are many such configurations, with particle numbers in the whole range between $N_{\Lambda}$ and $L_{\Lambda}$. So in the small staggering limit we find that there is a large ground state degeneracy. To reveal the nature of the ground states for small but non-zero $y$ we have to do degenerate perturbation theory. Remember that the degeneracy is lifted at second order in perturbation theory provided that the 'tic-tac-toe' lemma holds. We easily verify that this is indeed the case: since all elements of $\mathcal{H}_{Q_1}$ have the same number of particles on $S_1$, namely zero, we must have that $\mathcal{H}_Q = \mathcal{H}_{Q_2} (\mathcal{H}_{Q_1})$. It follows that the states with particle numbers between, but not equal to, $N_{\Lambda}$ and $L_{\Lambda}$, have an energy of order $y^2$. We can thus focus on the sectors with $N_{\Lambda}$ and $L_{\Lambda}$ particles. In the next sections we construct the effective Hamiltonians for these sector and show that the ground states are quantum liquids.

\subsection{Projected four-state clock model}
In this section we will construct the effective Hamiltonian and obtain an exact expression for the ground state in the sector with $N_{\Lambda}$ particles. For simplicity we first consider the case where the original lattice $\Lambda$ is the square lattice. As for the chain (see section \ref{sec:deglift}) we obtain the following general expression for the effective Hamiltonian that couples the states $\ket{\psi_k}$ that have zero energy for $y=0$:
\bea
H^{\text{eff}}_{lk} =  - \sum_m \frac{ \bra{\psi_l} H_1 \ket{\chi_m}\bra{\chi_m} H_1 \ket{\psi_k}}{E_m} + \bra{\psi_l}  H_2  \ket{\psi_{k}}.\nonumber
\eea
For $F=N_{\Lambda}$ the states $\ket{\psi_k}$ are such that all $S_1$ sites have precisely one occupied neighbor. It follows that each site can be in four states corresponding to the position of the occupied neighbor. We denote these four states $\ket{\sigma}$ as up, down, right and left: $\ket{U}$, $\ket{D}$, $\ket{R}$ and $\ket{L}$ (see figure \ref{fig:clockmap}). With this notation the effective Hamiltonian takes the simple form
\beq\label{eq:ccm}
H^{\text{eff},N_{\Lambda}}= - \sum_{i \in S_1}  \sum_{\sigma,\sigma'} \tilde{P} \ket{\sigma}\bra{\sigma'} \tilde{P} + \sum_{i \in S_2} P_{\langle i \rangle},
\eeq
where $\tilde{P}$ projects onto the states $\ket{\psi_k}$. The first term represents flipping the state of the vertex from one orientation to another. Due to the hard-core character of the fermions not all flips are allowed, for instance a vertex cannot flip into the up-state if the vertex above it is in the down-state. The projection operators $\tilde{P}$ impose these constraints. The second term is purely diagonal. The easiest way to compute it is by realising it counts the number of sites on $S_2$ whose neighbors are empty. We will discuss the second term in more detail when we consider the general case. For now we just mention that for the doubly decorated square lattice with doubly periodic boundary conditions this term is equal for all $\ket{\psi_k}$, so it is just an overall constant.

Remarkably, the first term in the Hamiltonian is precisely the Hamiltonian of the constrained clock model considered in \cite{Pielawa11} apart from an overall constant. In \cite{Pielawa11} the model arrises as an effective Hamiltonian for ultra cold bosons in a strongly tilted optical lattice. The lattice is a singly decorated square lattice which is tilted along the diagonal. In the limit of a strong tilt, the true ground state of this system simply has all bosons occupying the site in the lowest corner. However, if one starts from a Mott state (one boson per site) and adiabatically tilts the lattice the true ground state cannot be reached, because there is too much excess energy in the system. Instead the resonant subspace turns out to be presicely that of the constrained four-state clock model. It may be surprising that both a bosonic and a fermionic system map to the same effective model. This can be understood by noting that in the limit of $y \to 0$ the fermions get confined to the sites neighboring one vertex. Consequently, they cannot hop past eachother any more and effectively lose their fermionic character. 

The effective Hamiltonian, (\ref{eq:ccm}), is again of the Perron-Frobenius type, which ensures that the ground state is unique and nodeless. It was shown in \cite{Pielawa11} that the ground state of the constrained clock model with doubly periodic boundary conditions is the equal superposition of all states $\ket{\psi_k}$. This results carries over directly to the ground state of (\ref{eq:ccm}) on the doubly decorated square lattice with doubly periodic boundary conditions, since in that case the two models only differ by an overall constant. In the supersymmetric model the second term ensures that the ground state has zero energy. 

We now turn to the general case and show that, contrary to the constrained clock model, the second term also ensures that the equal superposition state is always the ground state independent of the original graph $\Lambda$ and the imposed boundary conditions \footnote{The only type of boundary we do not allow is the case where open boundary conditions are imposed by cutting links in the $\Lambda_3$ graph that connect sites of the different sublattices. As explained before, for this type of boundary condition the cohomology vanishes in the sector with $N_{\Lambda}$ fermions.}. To see this we define $\kappa_2$, which counts the number of links between two empty $S_2$ sites and $\kappa_1$ which counts the number of empty $S_2$ sites at the boundary of the graph. Note that $\kappa_1$ is nonzero only if we impose open boundary conditions by cutting links in the $\Lambda_3$ graph that connect two $S_2$ sites. Using this definition we can write the second term in the Hamiltonian as $N_{\Lambda}+2\kappa_2+\kappa_1$. We now proceed as in \cite{Pielawa11}, that is, we count the number of states to which the Hamiltonian connects some state $\ket{\psi_k}$. Note that the first term will change the state of the site by occupying a different neighboring $S_2$ site, however, this can only happen if the $S_2$ site has all neighboring sites empty. The number of such $S_2$ sites that can be occupied is precisely $2 \kappa_2 + \kappa_1$. Using this result we show that each row in the effective Hamiltonian sums to zero, which implies that the equal amplitude state has zero energy. We just showed that the off-diagonal elements in each row sum to $-2 \kappa_2 - \kappa_1$. One easily checks that the diagonal term is $- N_{\Lambda}+N_{\Lambda}+2\kappa_2+\kappa_1$. Clearly, the total sum is zero. Since we know that there is a unique zero energy state for $y>0$, we find that the equal amplitude state is the ground state. Note that this is also true if the sites in the graph $\Lambda$ have a different connectivity, this just means that there is a different number of clock states $\ket{\sigma}$. The fact that this holds for general $\Lambda_3$ graphs and boundary conditions is again a magical property of supersymmetry.

In \cite{Pielawa11} it is shown that correlation lengths in the ground state of the constrained clock model are finite, which suggests that the phase is gapped. This result carries over directly to the staggered supersymmetric model, where the original graph is the square lattice with doubly periodic boundary conditions. For the other cases, however, it is not clear whether the correlation lengths are still finite. Nevertheless, since the ground state is an equal superposition state, this can again be investigated using transfer matrices.

\begin{figure}
\centering
\includegraphics[width=0.6\textwidth]{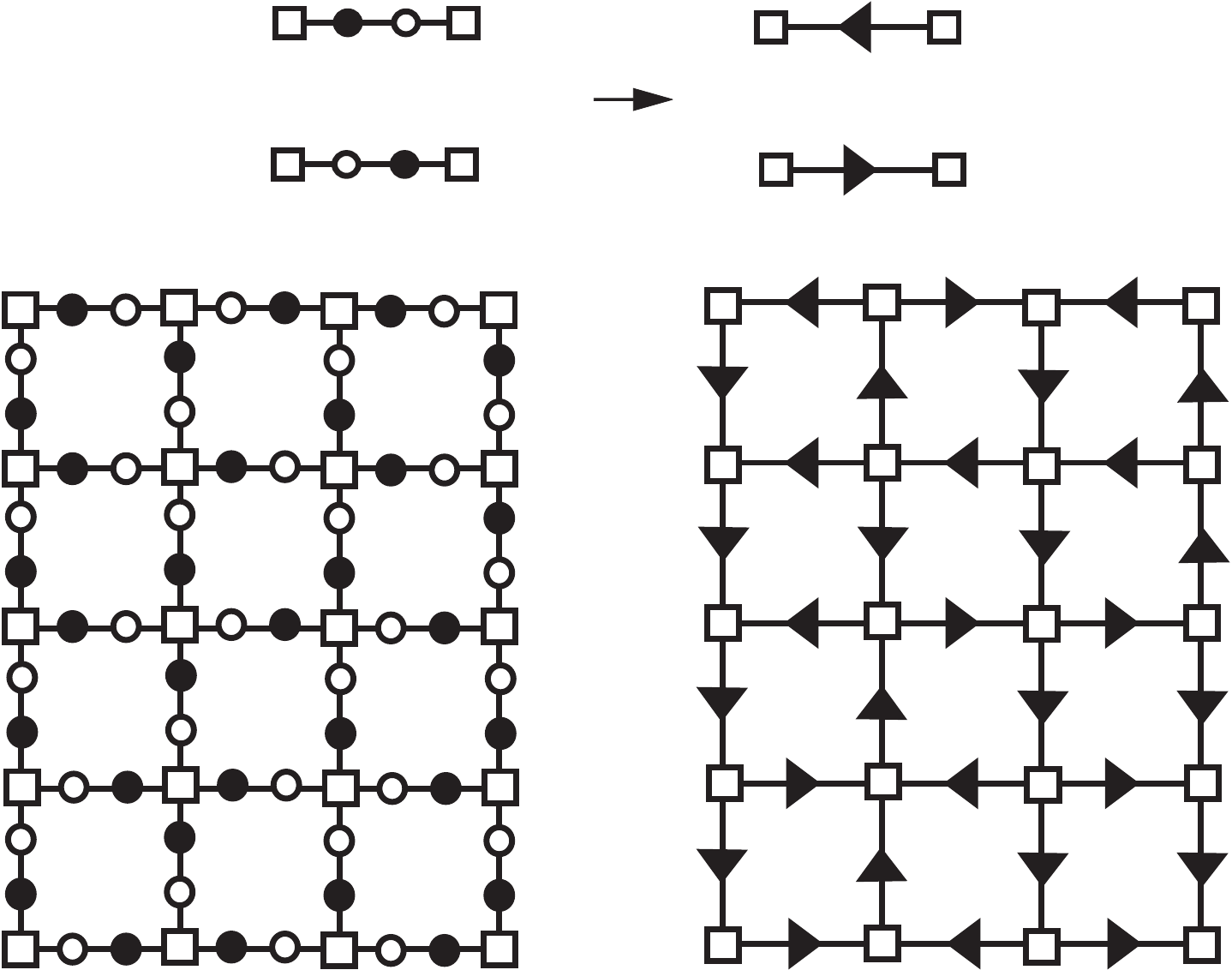}
\caption{Mapping of elements of the cohomology of $Q_1$ at $F=L_{\Lambda}$ to states on the original square lattice, where the degrees of freedom are arrows living on the links.}
\label{fig:arrowmap}
\end{figure}

\subsection{Projected two-state link model}
We now turn to the case with $F=L_{\Lambda}$. Remember that for $y \to \infty$ the ground state in this sector is the product state of two-site chain ground states. The ground state of a two site chain is the singlet $(c_1^{\dag}-c^{\dag}_2)/\sqrt{2}\ket{0}$. The product state thus has singlets resonating on the two $S_2$ sites that are added to each link of the original graph. We denote this state by $\vbs$ inspired by the obvious similarity with the valence bond solid states. We now show that in the limit of $y \to 0$ the ground state is the state $\tilde{P} \vbs$, where $\tilde{P}$ again projects onto the states $\ket{\psi_k}$, i.e. the cohomology elements of $Q_1$. Before we write the effective Hamiltonian for this sector, we map the configurations $\ket{\psi_k}$ to arrow configurations on the original graph $\Lambda$, where the arrows live on the links. The number of particles in this sector is precisely the number of links in $\Lambda$. We replace the two added sites by an arrow pointing towards the vertex which has an occupied neighbor on that link. The mapping is shown in figure \ref{fig:arrowmap}. It follows that all links have an arrow. Furthermore, each vertex has at least one arrow pointing inwards. If we now denote the two possible arrow orientations on a link by $\ket{\tau}$, we can write the effective Hamiltonian as
\beq
H^{\text{eff},L_{\Lambda}}= \sum_{\text{links}}  \sum_{\tau,\tau'} \tilde{P} \ket{\tau}\bra{\tau'} \tilde{P} - \sum_{\text{vertices}} \ket{1} \bra{1},
\eeq
where $\ket{1}$ denotes the vertex configuration with 1 arrow pointing in and the rest of the arrows pointing out. As before we count the number of configurations to which the Hamiltonian connects some state $\ket{\psi_k}$. Note that all arrows can be flipped except if that arrow is the only inpointing arrow at some vertex. So the Hamiltonian connects to $L_{\Lambda}-\nu_1$ states, where $\nu_1$ is the number of vertices with configuration $\ket{1}$ in the state $\ket{\psi_k}$. One easily checks that the diagonal term is precisely this number. To check that the state $\tilde{P} \vbs$ is a zero energy eigenstate of this Hamiltonian, we note that configurations which map into each other by flipping one arrow, have opposite signs in this state. From this one readily checks that $\tilde{P} \vbs$ is a zero energy eigenstate and thus the ground state in the limit of $y \to 0$.

The projected VBS state is also a quantum liquid state. This can be seen by absorbing the minus sign in the single particle singlet states into the definition of the creation operator of the fermions. If we define the fermion creation operators on each link as $f^{\dag}_1=c^{\dag}_1$ and $f^{\dag}_2=-c^{\dag}_2$, the single particle state reads $(f_1^{\dag}+f^{\dag}_2)/\sqrt{2}\ket{0}$ and the projected VBS state clearly is an equal amplitude superposition state. We note that the construction of this state is reminiscent of the construction of the AKLT state. In the case of the AKLT state, however, one projects out many more states per vertex than we do here.

\subsection{Summary of the different phases}
We conclude that we have found exact ground states in the limits for $y \to \infty$ and $y \to 0$ for the $\Lambda_3$ graphs. In this last section we discuss the character of the corresponding phases. In the limit of $y \to \infty$ the ground state at $F=N_{\Lambda}$ is a simple Wigner crystal with all sites of the original graph $\Lambda$ occupied, whereas the ground state at $F=L_{\Lambda}$ is the VBS state defined above, i.e. singlets on all added sites. It is clear that the correlation length is zero for the state with $N_{\Lambda}$ particles and at most two lattice spacings for the state with $L_{\Lambda}$ particles. In the other limit, we obtain two types of quantum liquid states. These states are equal superpositions of all states in the low energy subspace. For these states we do not know what the correlation length is in general. Both states can be obtained by a projection of single particle states onto the space of cohomology elements. This projection induces some correlation. However, intuitively one would not expect that this will induce a diverging correlation length in a two dimensional system, since it is just not constraining enough. This reasoning suggests that the spectrum in the limit of small $y$ is also gapped in both particle number sectors. As we will see in the next section, this is not true in one dimension. 

The ground states in the various gapped phases are quite different in character. In particular, for the case with $F=N_{\Lambda}$, where in one limit we have the liquid state and in the other limit the completely ordered state. Nevertheless, none of these states breaks any symmetry of the Hamiltonian. This means that there is no obvious order parameter we can write down that distinguishes the phases. It follows that it remains unclear whether or not there is a phase transition at intermediate values of $y$.

\section{Massless kink in staggered chain}\label{sec:kink}
In this section we return to the staggered chain discussed in section \ref{subsec:chaincohom}. It is clear that the staggered chain falls into the class of staggered $\Lambda_3$ graphs, where the original graph $\Lambda$ is also a chain. In section \ref{subsec:chaincohom}, we considered open boundary conditions imposed by cutting a link that connects two $S_2$ sites and found that there is a unique ground state. In the limit of large staggering it corresponds to the state with all $S_1$ sites occupied. In the limit of small staggering the ground state is the quantum liquid state: an equal superposition of all configurations in the low energy subspace. In the previous section we argued that for the two dimensional graphs all excitations over the quantum liquid state are gapped. Here, however, we will see that the liquid state supports a gapless mode. 

To appreciate the character of the gapless mode we briefly review some known results for the supersymmetric model on the chain. For $y=1$, the uniform chain, it has been shown that the system is integrable and the continuum limit is described by an $\mathcal{N}=(2,2)$ superconformal field theory with central charge $c=1$ \cite{fendley-2003-90,fendley-2005-95,Huijse11}. This indicates that the model is gapless for $y=1$. Furthermore, it was found that when periodic boundary conditions are imposed and the length of the chain is a multiple of three, there are two zero energy states. Staggering the Hamiltonian away from the critical point by setting $y\neq1$ has been identified with perturbing the continuum theory with the relevant supersymmetry preserving operator of conformal dimension 4/3 \cite{Fendley10a,Fendley10b}. For the chain with periodic boundary conditions and length a multiple of three it was shown that the system flows to a gapped phase both for $y>1$ and $y<1$. Furthermore, it was found that the ground state is twofold degenerate for all values of $y$, including $y=0$. Note that this is a special case, because we typically found that the ground state degeneracy of $\Lambda_3$ graphs is increased at $y$ strictly zero. We understand this by noting that on the periodic chain there are only two configurations that satisfy the condition that all sites of the original graph have at least one neighbor occupied:
\bea\label{eq:LRorder}
& &\bullet \square \circ \bullet \square \circ \dots \bullet \square \circ \bullet \square \circ, \nonumber\\
& &\circ \square \bullet \circ \square \bullet \dots \circ \square \bullet \circ \square \bullet.
\eea
Note that the two ground states are characterized by two types of order. By left-order we denote the phase in which the sites to the left of the $S_1$ sites are occupied and by right-order we denote the phase in which the sites to the right of the $S_1$ sites are occupied.

In this section we find that the gapless mode, supported by the open chain, corresponds to a kink separating the two types of order that characterize the two ground states on the periodic chain. Even though this kink is massless, the spectrum of the periodic system is gapped because the anti-kink is gapped. This is easy to see in the limit of small $y$; one cannot have the right-order to the left of the left-order in a state where all $S_1$ sites have at least one neighbor occupied. An anti-kink thus looks as follows
\bea
\dots \circ \square \bullet \circ \square \bullet \circ \square \circ \bullet \square \circ \bullet \square \circ \dots 
\eea
and costs an energy of order 1 in the small $y$ limit. In a periodic system one can only excite kink-anti-kink pairs, whereas in an open system one can have one without the other. This explains the somewhat counter intuitive fact that changing the boundary conditions changes the system from gapped to gapless. The way one should really think about this, is that by changing the boundary conditions one goes to a different sector. In one sector the number of kinks equals the number of anti-kinks and in the other sector there is one more kink than there are anti-kinks.

We will now analyse the staggered chain with open boundary conditions in more detail and show that the kink spectrum is similar to the spectrum of a particle in a box. We thus say that this is a one dimensional system which supports a zero dimensional massless mode.

Let us introduce the following notation. If we define
\bea\label{eq:defLR}
\ket{L_n^m} \equiv \prod_{k=n}^{m} c_{3k}^{\dagger}  \ket{\emptyset} \ \textrm{and} \  \ket{R_n^m} \equiv\prod_{k=n}^{m}c_{3k+2}^{\dagger}  \ket{\emptyset}  ,
\eea
we can write the states $\ket{\psi_i}$ and $\ket{\phi_{i}}$ defined in (\ref{eq:psik}) and (\ref{eq:phik}) as
\bea 
\ket{\psi_{i}} =\ket{L_0^{i-1}} \ket{R_{i}^{j-1}}  , \nonumber\\
\ket{\phi_{i}} =\ket{L_0^{i}} \ket{R_{i}^{j-1}}  . \nonumber
\eea
With this notation the two ground states of the periodic chain of length $L=3j$ read $\ket{L_0^{j-1}}$ and $\ket{R_0^{j-1}}$. It is clear that the states $\ket{\psi_i}$ and $\ket{\phi_i}$ are characterized by the location of a kink between the two types of ordering present in the two ground states of the periodic chain. 

As we have seen the ground state degeneracy at $y=0$ is lifted by the effective Hamiltonian, $W$ (\ref{eq:Heffchain}). It can easily be verified that this is simply the one dimensional version of the constrained clock model discussed in the previous section. The eigenvalues of $W$ are
\beq
\epsilon_m = 4 \sin^2 \left( \frac{m \pi}{2j+2} \right),
\label{eqn:low_lying_energies}
\eeq
with $0\leq m \leq j$. The corresponding wave functions are found to be
\beq\label{eq:pert_wavefns}
\sum_{k=0}^{j} \cos \left( \frac{\pi m (k+1/2)}{j+1} \right) \ket{\psi_{k}}.
\eeq
The ground state ($m=0$) is the state in which the kink between the two types of order is completely delocalized. Clearly, the kink spectrum is massless, because for large chain lengths $L=3j$ the energy of the low-lying states scales as $y^2/j^2$. Furthermore, we note that the ground state has long range correlations which are associated to the delocalization of the kink. To see this we note that in Wigner crystal type ordered states correlators such as $\langle n_0 n_{3i} \rangle - \langle n_0 \rangle \langle n_{3i} \rangle$ would be stricly zero. For example, in the states $\psi_j$ this correlator vanishes. However, after delocalization of the kink due to the gapless mode this correlator becomes non-zero. Indeed, we find for the normalized ground state (eqn (\ref{eq:pert_wavefns}) with $m=0$)
\beq
\langle n_0 n_{3i} \rangle - \langle n_0 \rangle \langle n_{3i} \rangle = \frac{j-i}{(j+1)^2}. \nonumber
\eeq
If we define $x=i/j$, and remembering that $j$ is proportional to $L$, we find that this correlator behaves as $(1-x)/L$. We associate the $1/L$ decay with the massless kink, which is essentially a zero dimensional mode. It is clear that this is in contrast with the exponential decay of correlators observed for the constrained clock model. The fact that we do not observe powerlaw decay with $x$ comes from the absence of truly one dimensional critical modes.

\begin{figure}
\centering
\includegraphics[width=0.75\textwidth]{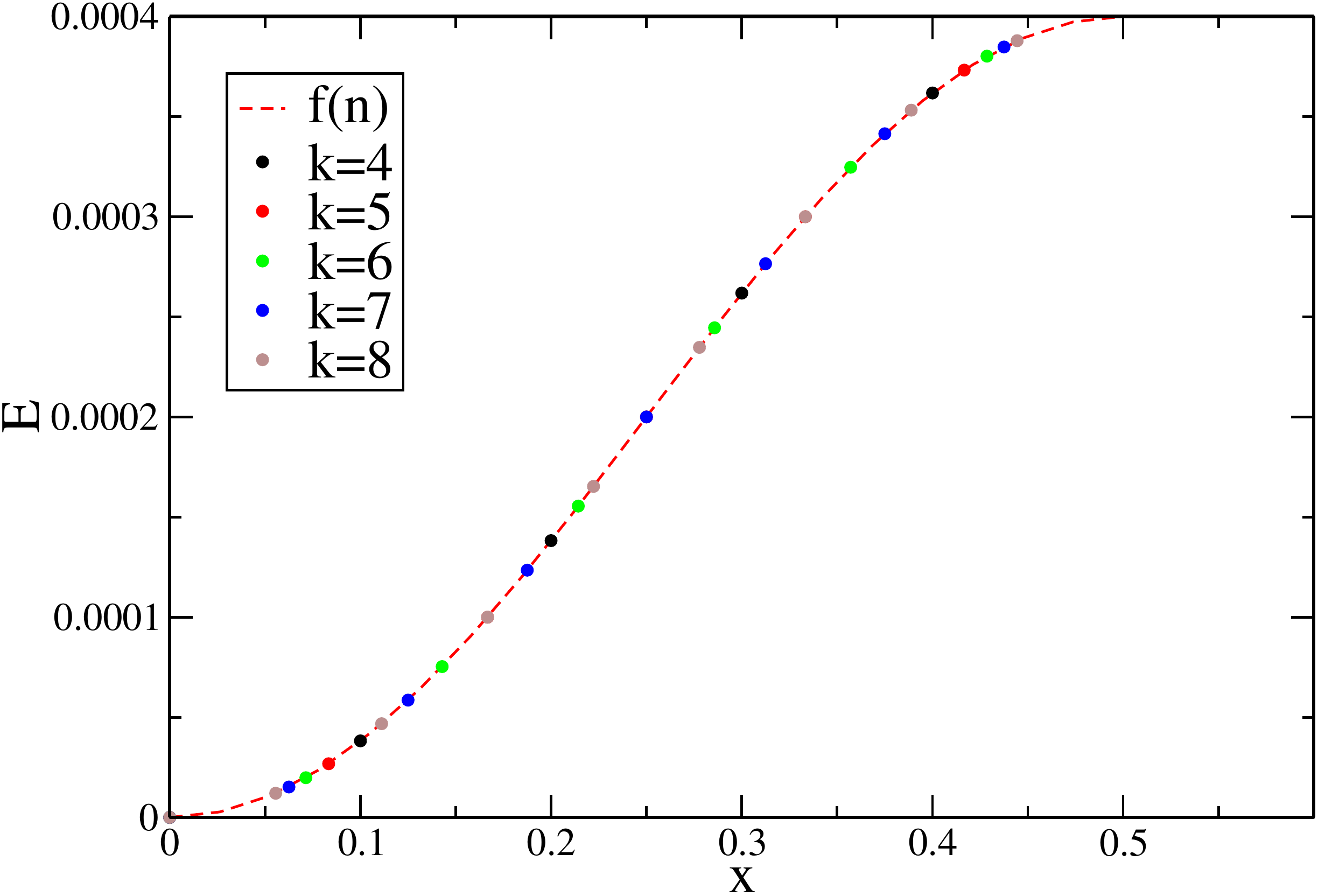}
\caption{Plot of the energies of low lying excitations of open staggered SUSY chains with staggering $y=1/100$ and lengths $3k$ with $4 \le k \le 8$. $x = \frac{n}{2k + 2}$. The equation used for the fit is $f(n) = 4 y^2 \sin^2(\frac{n \pi}{2k+2})$.}
\label{fig:low_lying_energies_plot}
\end{figure}

Figure \ref{fig:low_lying_energies_plot} shows a plot of the low lying energies for various chain lengths for $y=1/100$ alongside the expression in eqn (\ref{eqn:low_lying_energies}). Very good agreement is observed between the numerically calculated values and the prediction from the perturbative expansion. We have also verified an excellent agreement of the wave functions and their perturbative expression (\ref{eq:pert_wavefns}). Note that in the continuum limit both the energies and the wave functions are very similar to those of a particle in an infinite potential well. The differences are due to the fact that the kink experiences an attractive potential at the edges of the well. We say that the kink spectrum corresponds to a zero dimensional massless mode in a one dimensional system. This is in contrast to the massless, quantum critical phase observed for $y=1$, which is truly one dimensional.

In the following paragraphs we discuss the density and entanglement properties of the kink phase (see also \cite{Moran10}). The numerical results presented here and in the following are from calculations performed using large scale iterative exact diagonalisation techniques on large parallel distributed memory machines. The freely available code named DoQO (Diagonalisation of Quantum Observables) was employed for this purpose \cite{Moran11}.

\subsection{The ground state to higher order in $y$}
Now that we know the ground state to zeroth order we can go back to (\ref{eq:psi1coeff}) to obtain the ground state to first order in $y$. As we noted before $H_1$ only has non-zero elements between configurations which differ in their number of particles on the sites $3i+1$ by one. Since the $\ket{\psi_k}$ have no particles on these sites, $H_1$ maps them to configurations with one particle on this sublattice:
\bea\label{eq:gs1st}
H_1 \ket{\Psi_0} &=& H_1 \sum_{k=0}^{j} \ket{\psi_k} \nonumber\\
&=&  \sum_{k=0}^{j} \Bigg [ \sum_{p=0}^{k-1}  \ket{L_0^{p-1}} \otimes \ket{\circ \blacksquare \circ}_p \otimes  \ket{L_{p+1}^{k-1}} \otimes \ket{R_{k}^{j-1}}  \nonumber\\
& &+   \sum_{q=k}^{j-1}  \ket{L_{0}^{k-1}}\otimes \ket{R_{k}^{q-1}}  \otimes \ket{\circ \blacksquare \circ}_q \otimes   \ket{R_{q+1}^{j-1}} \Bigg ]\nonumber\\
&=&  \sum_{k=0}^{j} \Bigg [ \sum_{p=0}^{k-2}  \ket{L_0^{p-1}}  \otimes \ket{\circ \blacksquare \circ}_p \otimes   \ket{L_{p+1}^{k-1}} \otimes \ket{R_{k}^{j-1}}  \nonumber\\
& &+   \sum_{q=k+1}^{j-1}  \ket{L_{0}^{k-1}}\otimes  \ket{R_{k}^{q-1}}  \otimes \ket{\circ \blacksquare \circ}_q \otimes   \ket{R_{q+1}^{j-1}}  \Bigg ] \nonumber\\
& & + 2  \sum_{k=0}^{j-1}  \ket{L_{0}^{k-1}}   \otimes \ket{\circ \blacksquare \circ}_k \otimes   \ket{R_{k+1}^{j-1}}  .
\eea 
Here we introduced the notation
\bea\label{eq:def0blsq0}
\ket{\circ \blacksquare \circ}_k = c_{3k+1}^{\dag} \ket{\emptyset}.
\eea
It is easily checked that $H_0$ on all these configurations gives one. It follows that in  (\ref{eq:psi1coeff}) we have $E_m=1$ and we find
 \bea\label{eq:gs1stsym}
\ket{\Psi_1} = - H_1 \ket{\Psi_0} .
\eea 

To obtain the wave function to higher order is more subtle, because $\langle \Psi_0| \Psi_s\rangle$ will typically not be zero for $s>1$. This is nicely explained in \cite{Fendley10b}. The solution is to redefine the ground state as
\bea
\ket{\Psi}=  \sum_{k=1}^{j+1}  a_k (y) \ket{\psi_{k}} + y \ket{\tilde{\Psi}_1} + y^2 \ket{\tilde{\Psi}_2} + \dots
\eea
with $\langle \psi_k| \tilde{\Psi}_s\rangle=0$  for all $s>0$ and $1 \leq k\leq j+1$. As is explained in \cite{Fendley10b}, the coefficients $a_k(y)$ are polynomials in even powers of $y$. This is directly related to the fact that $H_1$ maps between states with even and odd number of particles on the sites $3i+1$. The coefficients of $y^2$ in the polynomials $a_k(y)$ follow from the conditions $\langle \psi_k | \Psi_4\rangle=0$. Consequently, one has to go to fourth order in perturbation theory to obtain these coefficients. Using numerics we compute $a_k(y)$ to second order for $j$ up to 7 and, for a convenient normalization, we find
\beq
a_1 = a_{j+1}= 1 + \mathcal{O}(y^4), \quad a_m =1 - y^2 + \mathcal{O}(y^4) \ \textrm{for $2 \leq m \leq j$}.
\eeq

\subsection{Density}
The density on site $m$ is given by the ground state expectation value of the number operator $\langle n_m \rangle$. It is easy to see that the density only contains even powers of the staggering parameter. We investigate the density distribution in the limit of small $y$.  To zeroth order the ground state is given in (\ref{eq:gs0th}). Clearly, to zeroth order the density on the sites $3i+1$ is zero. Furthermore, one finds that the density on the other sites depends linearly on the distance to the edge. For a chain of length $L=3j$ it is given by:
\bea
\langle n_m \rangle &=& \frac{3j-m}{3(j+1)} \ \textrm{for $m=3i$} \nn
\langle n_m \rangle &=& 0 \ \textrm{for $m=3i+1$} \nn
\langle n_m \rangle &=& \frac{m+1}{3(j+1)} \ \textrm{for $m=3i+2$.} \nonumber
\eea
The density to second order can be obtained from the ground state wave function to first order (\ref{eq:gs1st})-(\ref{eq:gs1stsym}). Upon inspection one finds:
\bea\label{eq:density_2nd_order}
\langle n_m \rangle &=& \frac{3j-m}{3(j+1)} -y^2\frac{(3j-m)(j-1)+6(j+1)}{3(j+1)^2} \ \textrm{for $m=3i$} \nn
\langle n_m \rangle &=& y^2\frac{j+3}{(j+1)}  \ \textrm{for $m=3i+1$} \\
\langle n_m \rangle &=& \frac{m+1}{3(j+1)} -y^2\frac{(m+1)(j-1)+6(j+1)}{3(j+1)^2} \ \textrm{for $m=3i+2$} .\nonumber
\eea
To get the fourth order correction to this result, one has to go to third order for the ground state. This was done numerically for chains of lengths up to $L=21$. From these results we infer that the density on the sites $3i+1$ to fourth order is:
\beq\label{eq:density_4th_order}
\langle n_m \rangle = 
\begin{cases}
y^2\frac{j+3}{(j+1) } - y^4\frac{j^2+8j+11}{(j+1)^2} & \quad \textrm{for $m=1, L-2$} \\
y^2\frac{j+3}{(j+1)} - y^4\frac{3j^2+10j+11}{(j+1)^2} &  \quad \textrm{for $m=3i+1$ and $1 \leq i \leq j-2$}. 
\end{cases}
\eeq

\begin{figure}
\centering
\subfloat[]{
\includegraphics[scale=0.5]{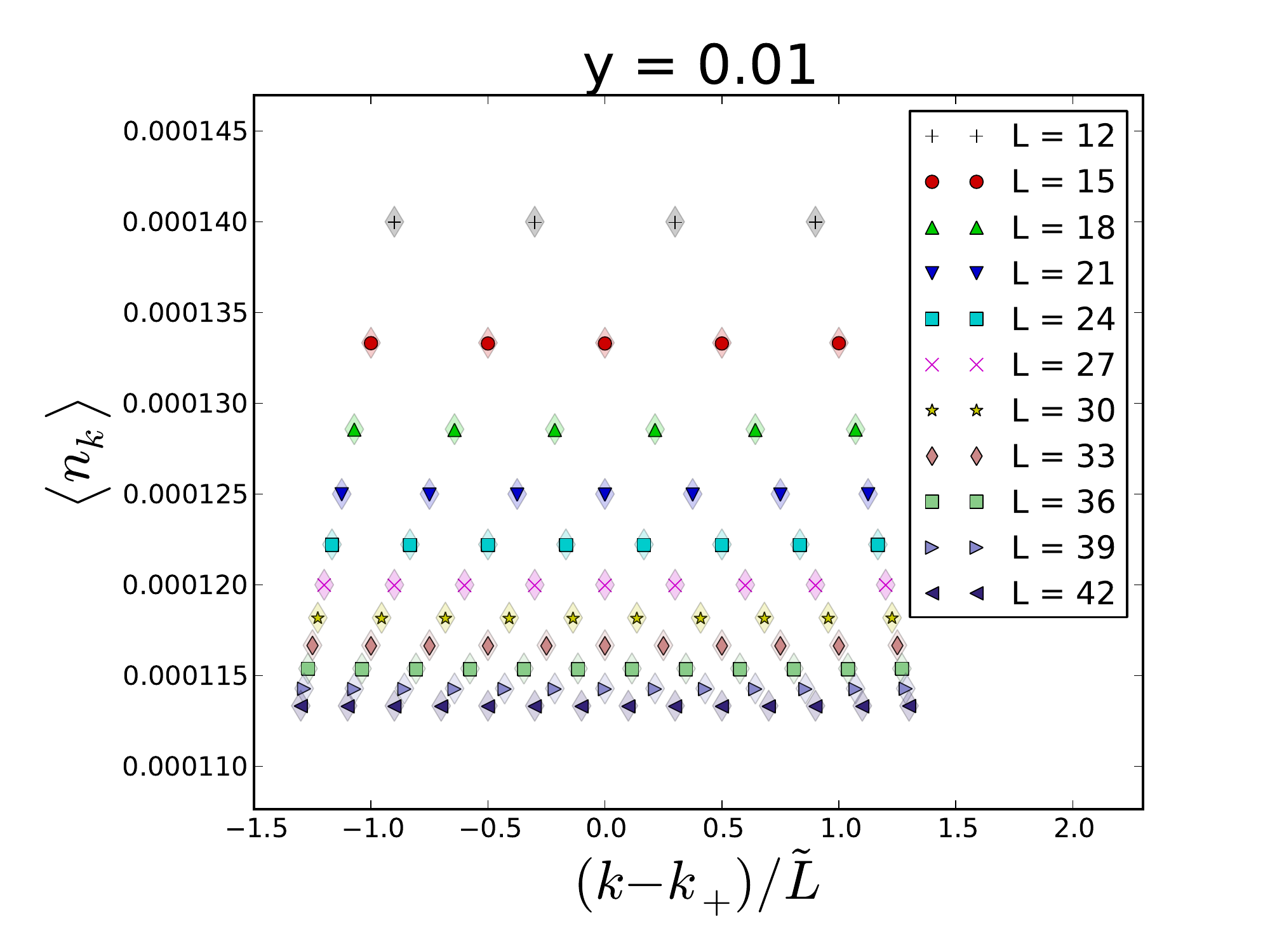}
}\\
\subfloat[]{
\includegraphics[scale=0.5]{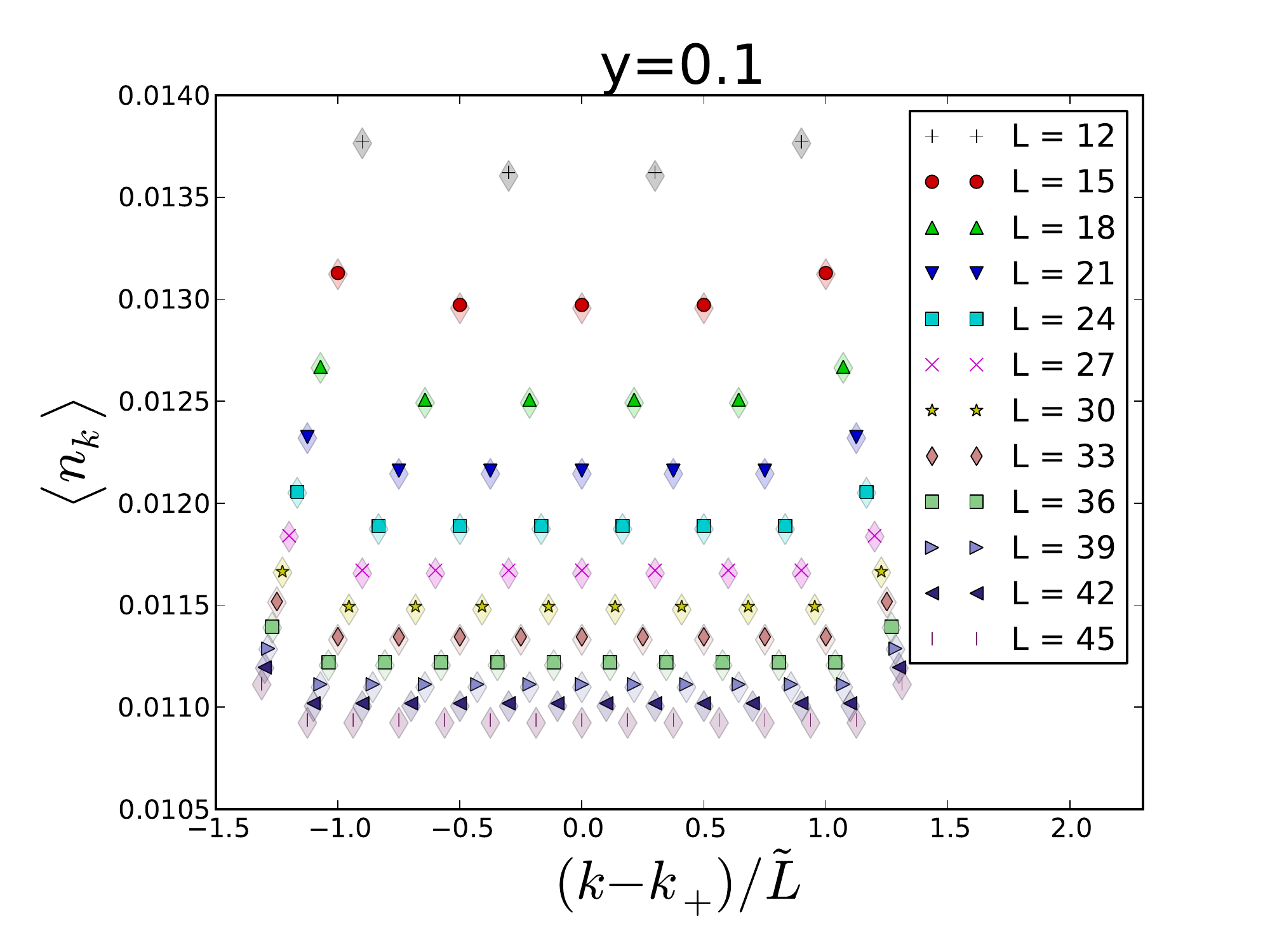}
}
\caption{Density on staggered sites $\langle n_i \rangle$ for $i \mod 3 = 1$ for (a) $y = 1/100$ and (b) $y = 1/10$. Shaded plotmarkers correspond to the predictions from perturbation theory. In (a) this is to first order using expressions in equation (\ref{eq:density_2nd_order}) and in (b) this is to fourth order using expressions in equation (\ref{eq:density_4th_order}). Here we follow the notation in \cite{Beccaria05} and use $k_{\pm} = (L \pm 1) / 2$ and $\tilde{L} = L/3 + 1$.}
\label{fig:den_plots}
\end{figure}

Figure \ref{fig:den_plots} shows numerically calculated values of the one point functions along side the values obtained from first and fourth order perturbation theory. We note that these results are quite different from the density results obtained for the staggered periodic chain \cite{Fendley10a,Fendley10b}. For the periodic chain, the density turns out to have the special property that it is scale free. This means that the first $2j$ terms in the series expansion of the density in $y$ are independent of the system size for systems of length $L=3j$. Clearly, this special property is absent for the density in the ground state of the open chain. This is not surprising since the kink state is very different from the ordered ground states of the periodic chain. In the limit of large $y$, where the ground state of the open chain is very similar to one of the two ground states of the periodic chain (both have all $S_1$ sites occupied), it is not so clear whether or not the scale free property of the density would survive on the open system. We have studied the staggered density numerically in this limit for the chain with open boundary conditions and did not find any scale free properties.

\subsection{Entanglement entropy}
We consider the entanglement entropy in the limit of small staggering parameter $y$. In this limit the ground state is given by the expression in equation \ref{eq:gs0th}. From this we determine the entanglement entropy and entanglement spectrum between the subsystems when the chain is partitioned. Any state $\ket{\psi}$ on the full system can be expressed as a sum $\ket{\psi} = \sum_{\alpha,\beta} W_{\alpha,\beta} \ket{\psi_A^{\alpha}} \ket{\psi_B^{\beta}}$ where $W_{\alpha,\beta}$ is called the weight matrix and $\ket{\psi_A^i}$ and $\ket{\psi_B^j}$ are orthonormal states spanning spaces $\mathbf{H}_A$ and $\mathbf{H}_B$ respectively. When an open chain with $3j$ sites is divided into subsets $A$ and $B$ containing $l_{A}=3l$ and $l_{B}=3(j-l)$ sites, respectively, the weight matrix $W_{ik}$ has $m$ rows and $n$ columns where $m= l+1, n= j-l+1$. The elements of this matrix are:
\beq\nonumber
W_{ik} = \frac{1}{\sqrt{j+1}} ( \delta_{0,i} + \delta_{n-1,k} - \delta_{0,i}\delta_{n-1,k})
\eeq
Multiplying this matrix by its transpose results in an $m \times m$ matrix which is the reduced density matrix of the subset $A$.
\beq \nonumber
\rho_A = WW^{T} = Tr_{B}(\rho_{AB}) 
\eeq
Likewise $\rho_B = W^{T}W$.
The matrix elements of $\rho_A$ are:
\beq\nonumber
 [\rho_A]_{i,k} = \begin{cases} 
 \frac{1}{j+1} &\text{for } i > 0 \text{ or } k > 0 \\
 \frac{n}{j+1} &\text{for } i = k = 0  
 \end{cases}
\eeq
This is a rank two matrix with eigenvalues given by: 
\beq\nonumber 
\lambda_{\pm} =  \frac{1}{j+1} \left( \frac{m+n-1}{2} \pm \frac{\sqrt{(m-n-1)^2+4(m-1)}}{2} \right) 
\eeq
and unnormalised eigenvectors of the form: 
\beq
\label{eqn:eigvectors_even}
\ket{x_{\pm}} = \ket{\frac{n-m+1}{2} \pm \frac{\sqrt{(m-n-1)^2+4(m-1)}}{2}, 1, \dots,1}
\eeq

When the chain is divided into subsets with lengths $l_{A} = 3l-1$ and $l_{B} = 3(j-l)+1$ the weight matrix has dimension $m \times n$ with $m = l+1, n=j-l+2$. The elements of this matrix are given by the expression:
\beq\nonumber
W_{ik} = \frac{1}{\sqrt{j+1}} ( \delta_{0,i} + \delta_{n-1,k} - 2\delta_{0,i}\delta_{n-1,k})
\eeq
The related reduced density matrix resulting from the multiplication of this weight matrix by its transpose again has dimension $m \times m$ but in this case has elements given by: 
\beq\nonumber
 [\rho_A]_{ik} = \begin{cases} 
 \frac{1}{j+1} &\text{for } i > 0 \text{ and } k > 0 \\
 \frac{n-1}{j+1} &\text{for } i = k = 0  
 \end{cases}
\eeq
This is again a rank two matrix but with eigenvalues given by: 
\beq\nonumber \lambda_{0} =  \frac{m-1}{j+1} = \frac{l}{j+1}, \quad \quad \lambda_{1} = \frac{n-1}{j+1}=\frac{j-l+1}{j+1} \eeq
and unnormalised eigenvectors of the form: 
\beq
\label{eqn:eigvectors_uneven}
\ket{x_{0}} = \ket{1, 0, \dots,0},\quad \quad  \ket{x_{1}} = \ket{0, 1,  \dots,1}
\eeq

\begin{figure}[htp]
\begin{center}
\subfloat{
\label{fig:susy_stag_open_L_39_EE_a_50_both}
\includegraphics*[scale=0.5]{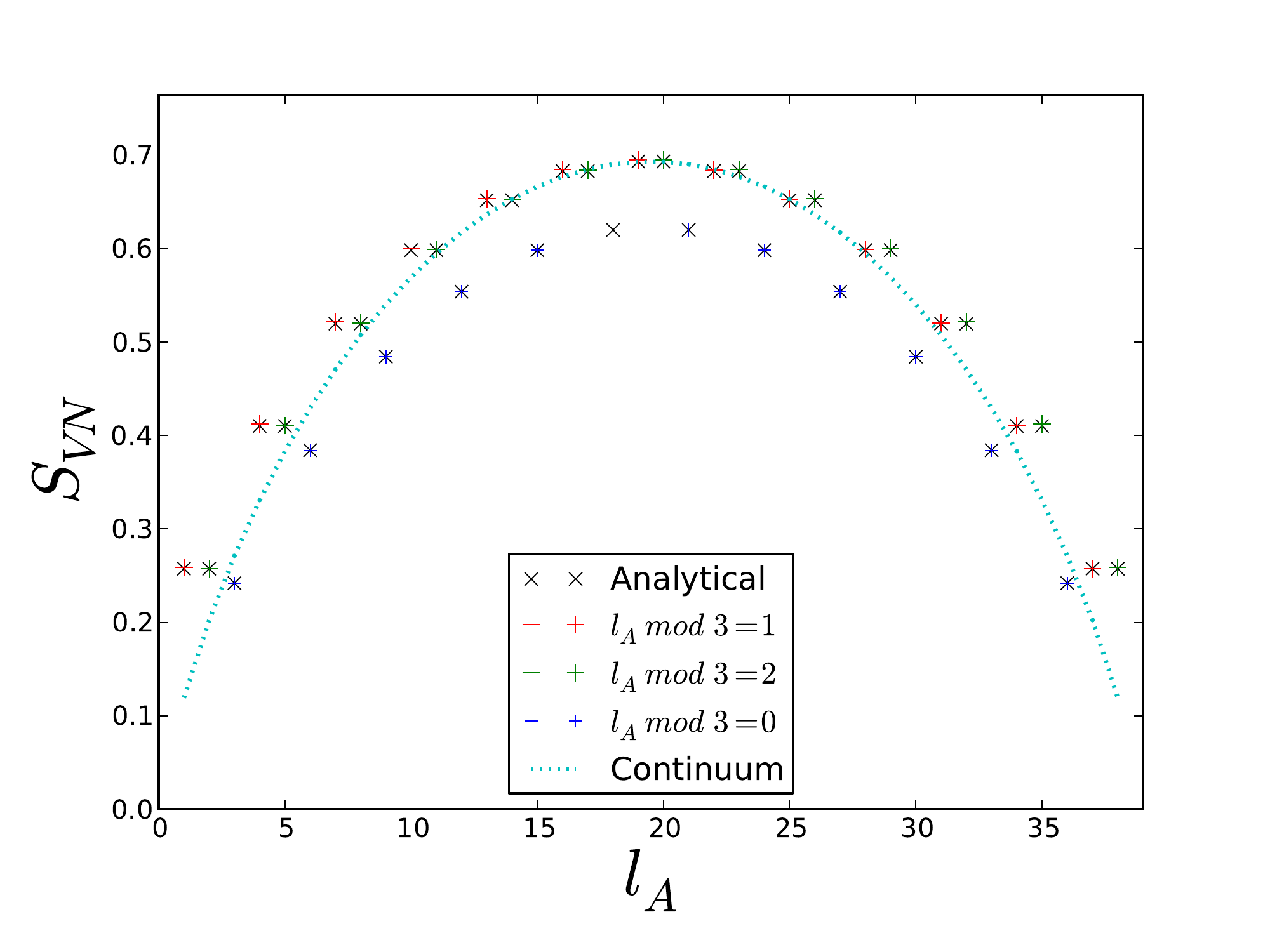}
}

\caption{Numerically calculated entanglement entropy data for the ground state of the open $L=39$ staggered chain with staggering parameter $y=\frac{1}{50}$ plotted alongside the analytical results for the small staggering limit indicated by the black crosses. The dotted line corresponds to the result for the system in the continuum limit where the $\mathbb{Z}_3$ substructure vanishes. As discussed in the text this line has a very simple interpretation in terms of the kink. 
}
\label{fig:open_stagg_chain_inf_stag_limit_ent}
\end{center}
\end{figure}

The entanglement entropy in terms of the eigenvalues of the reduced density matrix is $ S_{VN} = \sum_i -\lambda_i \log(\lambda_i) $ and the values of the entanglement spectrum are $ \xi_i = -\log(\lambda_i) $. Using the expression for the entanglement entropy the analytical results were compared to numerical data for an open chain with $L=39$ and small value of staggering parameter $y=\frac{1}{50}$ (figure \ref{fig:open_stagg_chain_inf_stag_limit_ent}). There is very good agreement between the analytical and numerical data.

In the limit of infinite system size where the system is partitioned into two equally sized subsets it can be seen from the expressions for the eigenvalues that both converge to 1/2. In this case the entanglement entropy is $S_{VN} = \log(2)$ and the entanglement spectrum consists of two points at $\log(2)$. 

The two points of the entanglement spectrum in each case can be interpreted in terms of the kink picture. In the case where the subset length is not a multiple of three, 
it is clear from the eigenvectors (\ref{eqn:eigvectors_uneven}) that the two eigenvalues of the reduced density matrix correspond to the cases where the `kink' is in subset A only or in subset B only. In the case where the chain is split into subsets with lengths that are multiples of three it is possible that as well as the kink being in either subset it can also be split by the partitioning such that half is in each subset. We define the vector $\ket{x_{A}}$ which corresponds to the kink being completely in subset $A$ and the vector $\ket{x_{\bar{A}}}$ which corresponds to the kink being either not in $A$ or split by the partitioning. In terms of these vectors the eigenvectors of $\rho_{A}$ (\ref{eqn:eigvectors_even}) can be expressed as: 
$\ket{x_{\pm}} = \frac{1}{\sqrt{2}}(\ket{x_{A}} \pm \ket{x_{\bar{A}}}) $.
In the infinite limit where the eigenvalues are degenerate, the superpositions $\frac{1}{\sqrt{2}}(\ket{x_{+}} \pm \ket{x_{-}})$ are also eigenvectors. Then $\frac{1}{\sqrt{2}}(\ket{x_{+}} + \ket{x_{-}}) = \ket{x_{A}}$ which corresponds to the kink being completely in $A$ and $\frac{1}{\sqrt{2}}(\ket{x_{+}} - \ket{x_{-}}) = \ket{x_{\bar{A}}}$ which corresponds to the situation where either the kink is not in $A$ or is split by the partitioning. 

Finally, we mention that the kink picture of the ground state allows us to derive an expression for the entanglement entropy as a function of subsystem size $l_A$ by writing a (trivial) wavefunction for the kink in the continuum limit. We note that this description does not capture the clear $\mathbb{Z}_3$ substructure present in the finite system. Since the kink is distributed uniformly over the chain, the continuum limit wavefunction for the kink is just $\Psi_{\textrm{kink}} (x)= 1/\sqrt{L}$ for $0<x<L$ and zero otherwise. Since the kink is either in part A or in part B, the reduced density matrix is
\beq\nonumber
\rho_A=P_A \ket{x_A}\bra{x_A} + P_{\overline{A}} \ket{x_{\overline{A}}}\bra{x_{\overline{A}}}.
\eeq
Where integrating out part B simply gives the probabilities of finding the kink in part A or part B, which follow from the wavefunction above: $P_A = l_A/L$ and $P_{\overline{A}}= (l_A-L)/L$. Plugging this into the formula for the entanglement entropy, one finds:
\beq\label{eq:EE_cont}
S_{VN} (l_A) = -l_A/L \log [l_A/L ] - (L-l_A)/L \log [(L-l_A)/L].
\eeq
The fact that in this case the levels of the entanglement spectrum can be so directly related to a physical feature of the system, namely the kink is a very interesting feature of this model. 

\section{Outlook}
An obvious follow-up on this work is to study the staggered supersymmetric model on other lattices, in particular the ones that exhibit its key feature: superfrustration. In \cite{fendley-2005-95} the limits of small and large staggering were considered for the supersymmetric model on the martini lattice. The martini lattice has an exponential number of ground states and the ground states are in one-to-one correspondence with dimer coverings of the underlying honeycomb lattice. In one limit of the staggering parameter the dimer coverings become exact ground states. It was shown that by introducing another parameter that breaks supersymmetry, this ground state degeneracy is lifted and the effective Hamiltonian is a quantum dimer model. 

We will briefly discuss two systems that seem particularly promising for continuing the strategy introduced in this paper in which supersymmetry is preserved: the square lattice decorated by one extra site on each link and the octagon-square lattice, which is obtained from the square lattice by replacing each vertex by a smaller square that is tilted by 45 degrees. When wrapped around a torus, both lattices show an exponential ground state degeneracy. Still the frustration is suppressed, since the ground state entropy is subextensive: it grows with the linear size of the system, not the area. 

Let us first consider the octagon-square lattice. The cohomology problem was solved in \cite{fendley-2005-95}. Interestingly, the cohomology problem can be solved for various choices of the two sublattices $S_1$ and $S_2$. Clearly, this makes this an interesting system to look at, since various staggerings can be considered. Let us briefly discuss a choice that seems particularly promising. Let the top site of each added square be in $S_1$, the $S_2$ is a collection of chains with length a multiple of three. If periodic boundary conditions are imposed in the horizontal direction, that is in the direction along the chains, the elements of $\mathcal{H}_{Q_2}$ have all $S_1$ sites empty. Since each periodic chain has two ground states, the number of elements is $2^{N_c}$, where $N_c$ is the number of chains. On the other hand, the elements of $\mathcal{H}_{Q_1}$ also have all $S_1$ sites empty, however, now the particles on the $S_2$ sites, must be arranged such that all $S_1$ sites have an occupied neighbor. It follows that all chains can be in one of the two configurations depicted in (\ref{eq:LRorder}). So again the number of elements in $2^{N_c}$. So even at strictly zero staggering, the number of zero energy states is unchanged. It follows that one cannot write an effective Hamiltonian that couples these states. What is interesting, however, is that for very large and very small staggering, this system behaves as a collection of one dimensional systems, which in one limit are quantum critical and in the other limit are gapped. If you wish, there is some resemblance with the one dimensional chain, where we had a zero dimensional gapless mode in a 1D system, whereas here we have one dimensional gapless modes in a two dimensional system. The question is of course what happens to the criticality when the staggering parameter is finite. Since the critical chains are sensitive to a relevant perturbation that drives them to the gapped phase that we observe in the limit of very small staggering, a possibility is that the system is gapped for all finite values of the staggering parameter. In that
case the question arises as to whether these $2^{N_c}$ ground states are topologically distinct or not.

Another interesting system to look at is the square lattice with one extra site on each link, which we will call the singly decorated square lattice. This lattice falls into the class of $\Lambda_2$ graphs. As we mentioned it can be proven \cite{Csorba09,HuijseT10} that the cohomology problem of these graphs maps to the cohomology problem of the original graph via a particle-hole inversion. The cohomology problem of the square lattice with doubly periodic boundary conditions has been solved \cite{Huijse10}. The ground states are found to have densities $\nu=F/N$, where $N$ is the total number of sites in the whole range between 1/5 and 1/4. For the singly decorated square lattice this means that there are zero energy states for all rational densities in the range [1/4, 4/15]. This is true for all non-zero values of the staggering parameter. As before we choose the sublattice $S_1$ to be the original lattice and $S_2$ to be the added sites. For the staggering parameter strictly equal to zero, all cohomology elements of $\mathcal{H}_{Q_1}$ are zero energy states. It turns out that these states can be found for all densities in the range [1/6, 2/3]. At 1/6 density, the cohomology elements are one-to-one with close packed dimer configurations on the links of the original square lattice. For any non-zero staggering we know that there are no zero energy states at 1/6 filling. Interestingly, we find that to second order the effective Hamiltonian in this sector is just a constant, so the states gain some finite energy as they should, but they remain degenerate. This degeneracy is lifted at fourth order, where the effective Hamiltonian is the usual plaquette flip term in the Rokhsar-Kivelson model \cite{Rokhsar88}. Since the potential term is absent (it arises only at eighth order) the system is in the gapped phase which shows columnar ordering of the dimers. The study of the effective Hamiltonians in the sectors at higher densities is beyond the scope of this project. From the cohomology structure is it clear that they may be quite involved, but therefore possibly also quite interesting. 

\section{Conclusion}
We showed that exact ground states for the supersymmetric model on decorated graphs can be obtained in the limit of small and large staggering. In one limit we find that the ground states are either a Wigner crystal or a simple VBS, in the other limit the ground states are quantum liquid states arising from a projection of a product state with one particle per plaquette (for $F=N_{\Lambda}$) or link (for $F=L_{\Lambda}$), reminiscent of the AKLT quantum liquid states. It turns out that the effective Hamiltonian in the sector with $F=N_{\Lambda}$ for the decorated square lattice with periodic boundary conditions is precisely the constrained clock model of \cite{Pielawa11}. Since the ground state is an equal superposition of classical configurations techniques from classical statistical mechanics could be used to study correlation functions in this state. They found an exponential decay for all correlation functions considered. This suggests that the system is gapped. Intuitively, we expect this to hold for all liquid states that arise on the $\Lambda_3$ lattices. The liquid state on the chain forms an exception since it does show long range correlations. The massless mode in this system is understood as a kink separating two types of order.  Apart from studying correlation functions, it would be interesting to look at the entanglement entropy of the liquid states, since a saturating entanglement entropy in a finite range of the subsystem size is a clear signature of a gap. 

The question of whether or not there is a phase transition for intermediate values of the staggering parameter can be addressed by numerical studies of ladder systems. The chain is clearly a very special case, because of the twofold degeneracy of the ground state when periodic boundary conditions are imposed. Decorated ladders, however, are expected to behave very similar to full two dimensional systems. 

Finally, a promising follow-up on this work would be to look at other lattices for which the cohomology problem can be solved in two limits of the staggering parameter. As we mentioned, the supersymmetric model typically has a vast ground state degeneracy, which will make this analysis more challenging. However, it also holds the promise of discovering exotic phases with frustration.

\section*{Acknowledgements}
LH would like to thank E. Berg and S. Pielawa for discussions and acknowledges financial support from the Netherlands Organisation for Scientific Research (NWO). KjS thanks B. Nienhuis for discussions on the integrability of the staggered chain model. JV and NM would like to thank the Science Foundation Ireland for support through the awards 05/YI2/I680 and 10/IN.1/I3013 and acknowledge the SFI/HEA Irish Centre for High-End Computing (ICHEC) for the provision of computational facilities and support.

\bibliography{literature}

\end{document}